\documentclass{emulateapj}
\usepackage{apjfonts}

\usepackage{rotating}

 \newcommand{\eg}{\emph{e.g.}}
 \newcommand{\kms}{\mbox{km\ \ensuremath{\rm{s}^{-1}}}}
 
\newcommand{\be}{\begin{equation}}
\newcommand{\ee}{\end{equation}}

\newcommand{\vycn}{C$_2$H$_3$CN}
\newcommand{\etcn}{C$_2$H$_5$CN}
\newcommand{\cyano}{HC$_3$N}
\newcommand{\mecn}{CH$_3$CN}

\usepackage{graphicx}	% Including figure files

\submitted{AJ; submitted March 2019; accepted for publication June 2019}

\begin{document}

\title{ALMA Spectral Imaging of Titan Contemporaneous with Cassini's Grand Finale}

\author{M. A. Cordiner\altaffilmark{1,2}, N. A. Teanby\altaffilmark{3}, C. A. Nixon\altaffilmark{1}, V. Vuitton\altaffilmark{4}, A. E. Thelen\altaffilmark{1,2}, S. B. Charnley\altaffilmark{1}}
%\author{M. A. Cordiner et al.\altaffilmark{1,2}}

\altaffiltext{1}{NASA Goddard Space Flight Center, 8800 Greenbelt Road, MD 20771, USA.}
\email{martin.cordiner@nasa.gov}
\altaffiltext{2}{Institute for Astrophysics and Computational Sciences, The Catholic University of America, Washington, DC 20064, USA.}
\altaffiltext{3}{School of Earth Sciences, University of Bristol, Wills Memorial Building, Queens Road, Bristol, BS8 1RJ, UK.}
\altaffiltext{4}{Universit{\'e} Grenoble Alpes, CNRS, IPAG, F-38000 Grenoble, France}

\begin{abstract}
The Cassini mission performed 127 targeted flybys of Titan during its 13-year mission to Saturn, culminating in the Grand Finale between April-September 2017. Here we demonstrate the use of the Atacama Large Millimeter/submillimeter Array (ALMA) to continue Cassini's legacy for chemical and climatological studies of Titan's atmosphere. Whole-hemisphere, interferometric spectral maps of HCN, HNC, \cyano, \mecn, \vycn, \etcn\ and C$_3$H$_8$ were obtained {using ALMA} in May 2017 at moderate ($\approx0.2''$, or $\approx1300$~km) spatial resolution, revealing the effects of seasonally-variable chemistry and dynamics on {the distribution of} each species. {The ALMA sub-mm observations of HCN and HC$_3$N are consistent with Cassini infrared data on these species, obtained in the same month.} Chemical/dynamical lifetimes of a few years are inferred for \vycn\ and \etcn, in reasonably close agreement with the latest chemical models incorporating sticking of \etcn\ to stratospheric aerosol particles. ALMA radial limb flux profiles provide column density information as a function of altitude, revealing maximum abundances in the thermosphere (above 600 km) for HCN, HNC, HC$_3$N and \etcn. This constitutes the first detailed measurement of the spatial distribution of HNC, which is found to be confined predominantly to altitudes above $730\pm60$~km. The HNC emission map shows an east-west hemispheric asymmetry of $(13\pm3)$\%. These results are consistent with very rapid production (and loss) of HNC {in Titan's uppermost atmosphere}, making this molecule an effective probe of short-timescale (diurnal) ionospheric processes.

\end{abstract}


%%%%%%%%%%%%%%%%%%%%%%%%%%%%%%%%%%%%%%%%%%%%%%%%%%

%%%%%%%%%%%%%%%%% BODY OF PAPER %%%%%%%%%%%%%%%%%%

\section{Introduction}

Saturn's largest moon Titan has a thick (1.45~bar) atmosphere composed primarily of molecular nitrogen ($98$\%) and methane ($\sim2$\%). Remote and in-situ measurements have identified a wealth of other species therein \citep[see][]{bez14}, the abundances, distributions and temperatures of which are linked to seasonal variations in Titan's combined photochemistry and atmospheric circulation patterns (\eg\ \citealt{tea09}; \citealt{vin10}; \citealt{ach11}; \citealt{cor15}; \citealt{the19}). Hydrocarbons and nitriles spanning a broad range in size and complexity have been detected, from acetylene (C$_2$H$_2$) and hydrogen cyanide (HCN) to benzene (C$_6$H$_6$), propionitrile (CH$_2$CH$_3$CN) and organic aerosols \citep{hor17}, which are theorized to originate from photochemistry initiated primarily by N$_2$ and CH$_4$ photolysis in the upper atmosphere \citep{vui19}. However, despite 13 years of dedicated study by the Cassini spacecraft during over 120 scientific flybys (from 2004-2017), major gaps remain in our knowledge of Titan's atmospheric chemistry and dynamics \citep{hor17,nix18}, particularly regarding the sources and sinks of complex organic molecules. 

Strong latitudinal anisotropies in the abundances of Titan's smaller hydrocarbons and nitriles (such as HCN, HC$_3$N and C$_2$H$_2$) were first observed by Voyager 1 \citep{cou95}. Continued monitoring of these, and other species, by the Cassini Composite Infrared Spectrometer (CIRS) has revealed temporal variations in the abundance patterns of photochemically-produced gases over time periods as short as (Earth) years or even months \citep{tea12,cou13,vin15,tea17}. The observed changes in molecular distributions are unique among the bodies of the Solar System, and are believed to be driven by seasonal insolation variations, mediated by a global meridional circulation cell that changes direction during Titan's 29.5 yr seasonal cycle. To provide new insights into Titan's atmospheric chemistry and global circulation system, we undertook a program to perform comprehensive mapping of short-lived photochemical products using the Atacama Large Millimeter/submillimeter Array (ALMA) and {Cassini CIRS} during the transition to Titan's northern summer solstice in May 2017. These observations capitalize on (1) the recent completion of the full array of $50\times12$~m ALMA antennas, enabling the highest-resolution studies of Titan to-date at mm/sub-mm wavelengths {(see also \citealt{lel19})}, and (2) the final opportunity for contemporaneous observations using ALMA and CIRS. The aim of this study was to combine the complementary diagnostic power of the sub-mm and infrared spectral regions to provide the most complete picture to-date of Titan's rapidly evolving chemistry and dynamics.

Cassini CIRS spectroscopy benefits from high sensitivity and spatial resolution for measuring rovibrational molecular emission lines, but provides an incomplete picture of Titan, hindered by a relatively low spectral resolution ($R\sim2000$) and reduced sensitivity at far-infrared wavelengths. Moreover, CIRS is only sensitive to a subset of Titan's gases (including CO, HCN, H$_2$O, HC$_3$N, C$_2$N$_2$ and various C$_n$H$_m$ hydrocarbons), and crucially, lacks the ability to detect several important and abundant nitrogen-bearing species. Polar nitriles (such as HNC, HC$_3$N, C$_2$H$_5$CN and C$_2$H$_3$CN) are among the species with the shortest predicted chemical lifetimes (less than a few years in the stratosphere; see \citealt{wil04,kra09,vui19}), and are therefore extremely useful probes of Titan's short-timescale atmospheric variability. The mm/sub-mm provides optimal sensitivity for rotational emission from polar nitriles, and with its combined high sensitivity and spatial resolution, ALMA can provide detailed maps of the emission from these species in Titan's atmosphere. Through radiative transfer analysis, the combined Cassini plus ALMA dataset enables molecular abundances to be derived as a function of spatial coordinate, from which a comprehensive evaluation of our understanding of the dominant nitrile formation and destruction pathways in nitrogen and carbon-rich planetary atmospheres can be obtained. Such information is crucial for constraining theories regarding the formation of aerosols and complex nitrogen-bearing organics (including possible pre-biotic species) in primitive (exo-)planetary atmospheres throughout the Galaxy.

The observations presented here provide an overview of the complete ALMA dataset obtained in synergy with the final Titan flybys by Cassini. Molecular maps for HCN, HNC, \cyano, \mecn, \vycn, \etcn\ and C$_3$H$_8$ demonstrate ALMA's unique, high-sensitivity interferometric imaging capabilities, resulting in spatially resolved distributions for individual molecules, from which we derive preliminary abundances and chemical/dynamical decay timescales. By comparison with results from the latest atmospheric chemical models, new insights are obtained into the relationship between Titan's climate, chemistry, and global atmospheric circulation system.

\section{ALMA Observations}
\label{sec:obs}

Observations of Titan were carried out using the ALMA Band 7 receiver on 2017-05-08 and 2017-05-16, as part of the director's discretionary time program 2016.A.00014.S. The observations were targeted to be close in time to Cassini's May 24 (solstice) flyby of Titan, for synergistic observations with CIRS. The ALMA correlator was configured using two different spectral settings to observe multiple frequency windows covering our molecules of interest in the range (1) 342-356~GHz (877-842~$\mu$m) and (2) 349-364~GHz (859-824~$\mu$m), at a spectral resolution of 244-977~kHz (0.2-0.8~\kms). For more details of the targeted spectral lines, see Table \ref{tab:spec}. 

\begin{table}
\centering
\caption{ALMA detected emission line frequencies, transitions and upper-state energies 
\label{tab:spec}}
\vspace{3mm}
{\footnotesize
\begin{tabular}{lllrr}
\hline\hline
Species          &Rest Freq.           & Transition              & Spec. Res.& $E_u$\\
                 &(MHz)                &                         & (kHz)     &  (K)  \\
\hline
CO               & 345796              & $3-2$                   &  977       & 33.2\\
HCN              & 354505              & $4-3$                   &  977       & 42.5\\
HCN              & 354460              & $4-3,\,v_2=1e$          &  977       & 1067\\
HNC              & 362630              & $4-3$                   &  244       & 42.5\\
H$^{13}$CN       & 345340              & $4-3$                   &  977       & 41.4\\
H$^{13}$CN       & 345239              & $4-3,\,v_2=1e$          &  977       & 1057\\
HC$^{15}$N       & 344200              & $4-3$                   &  977       & 41.3\\
HC$_3$N          & 345609              & $38-37$                 &  977       & 324\\
HC$_3$N          & 354697              & $39-38$                 &  977       & 341\\
HC$_3$N          & 355278              & $39-38,\,v_6=1e$        &  977       & 1059\\
HC$_3$N          & 355557              & $39-38,\,v_6=1f$        &  977       & 1059\\
HC$_3$N          & 355566              & $39-38,\,v_7=1e$        &  977       & 662\\
HC$_3$N          & 356072              & $39-38,\,v_7=1f$        &  977       & 662\\
HC$_3$N          & 363785              & $40-39$                 &  244       & 358\\
H$^{13}$CCCN     & 343739              & $39-38$                 &  977       & 330\\
H$^{13}$CCCN     & 361353              & $41-40$                 &  488       & 364\\
HC$^{13}$CCN     & 344143              & $38-37$                 &  977       & 322\\
HC$_3${$^{15}$N}   & 344385              & $39-38$                 &  977       & 331\\
CH$_3$CN         & 349025              & $19_8-18_8$             &  488       & 624\\
CH$_3$CN         & 349125              & $19_7-18_7$             &  488       & 517\\
CH$_3$CN         & 349212              & $19_6-18_6$             &  488       & 425\\
CH$_3$CN         & 349286              & $19_5-18_5$             &  488       & 346\\
CH$_3$CN         & 349346              & $19_4-18_4$             &  488       & 282\\
CH$_3$CN         & 349393              & $19_3-18_3$             &  488       & 232\\
CH$_3$CN         & 349427              & $19_2-18_2$             &  488       & 196\\
CH$_3$CN         & 349447              & $19_1-18_1$             &  488       & 175\\
CH$_3$CN         & 349454              & $19_0-18_0$             &  488       & 168\\
CH$_3${$^{13}$CN}  & 349221              & $19_3-18_3$             &  488       & 232\\
CH$_3${$^{13}$CN}  & 349254              & $19_2-18_2$             &  488       & 196\\
CH$_3${$^{13}$CN}  & 349274              & $19_1-18_1$             &  488       & 175\\
CH$_3${$^{13}$CN}  & 349280              & $19_0-18_0$             &  488       & 168\\
C$_2$H$_3$CN     & 360876              & $38_{8,31}-37_{8,30}$   &  488       & 476\\
C$_2$H$_3$CN     & 360876              & $38_{8,30}-37_{8,29}$   &  488       & 476\\
C$_2$H$_3$CN     & 360878              & $38_{9,29}-37_{9,28}$   &  488       & 512\\
C$_2$H$_3$CN     & 360878              & $38_{9,30}-37_{9,29}$   &  488       & 512\\
C$_2$H$_3$CN     & 361652              & $38_{4,35}-37_{4,34}$   &  488       & 373\\
C$_2$H$_5$CN     & 343195              & $38_{ 5,33}-37_{ 5,32}$ &  977       & 347\\
C$_2$H$_5$CN     & 354477              & $40_{ 3,38}-39_{ 3,37}$ &  977       & 361\\
C$_2$H$_5$CN     & 355756              & $39_{ 3,36}-38_{ 3,35}$ &  977       & 351\\
C$_2$H$_5$CN     & 349443              & $39_{10,30}-38_{10,29}$ &  488       & 446\\
C$_2$H$_5$CN     & 349443              & $39_{10,29}-38_{10,28}$ &  488       & 446\\
C$_2$H$_5$CN     & 349547              & $39_{ 9,31}-38_{ 9,30}$ &  488       & 425\\
C$_2$H$_5$CN     & 349547              & $39_{ 9,30}-38_{ 9,29}$ &  488       & 425\\
C$_2$H$_5$CN     & 349731              & $39_{ 8,32}-38_{ 8,31}$ &  488       & 407\\
C$_2$H$_5$CN     & 349731              & $39_{ 8,31}-38_{ 8,30}$ &  488       & 407\\
C$_2$H$_5$CN     & 349796              & $39_{ 4,36}-38_{ 4,35}$ &  488       & 354\\
C$_3$H$_8$       & 360978              & $8_6-7_5$               &  488       & 64.2\\
\hline
\end{tabular}
}
\parbox{\columnwidth}
{\footnotesize \vspace*{1mm} {\bf Notes.} Transitions are expressed as $J'-J''$, $J'_{K'}-J''_{K''}$ or $J'_{{K_a'},{K_c'}}-J''_{{K_a''},{K_c''}}$. C$_3$H$_8$ $8_6-7_5$ line is a blend of eight $\Delta K_a, \Delta K_c = 1$ transitions. Primary spectroscopic sources 
for molecular line frequencies: C$_2$H$_3$CN --- \citet{kis09}, C$_3$H$_8$ --- \citet{dro06}, HC$_3$N --- \citet{tho00}, C$_2$H$_5$CN --- \citet{bra09}, CH$_3$CN --- \citet{bou80}.}
\end{table}

The ALMA configuration was moderately extended, with 46 active antennas providing baselines in the range 15-1124~m, resulting in a spatial resolution of $\approx0.2''$ (using natural visibility weighting). The total on-source observing time was 18~min in setting 1 and 138~min in setting 2, leading to RMS noise levels in the range 2-4~mJy\,beam$^{-1}$\,MHz$^{-1}$. Weather conditions were very good, with zenith precipitable water vapor (PWV) in the range 0.74-0.83~mm, resulting in low atmospheric opacity and good phase stability.

After the initial calibration of the antenna phase delays, the observations consisted of bandpass and flux calibration scans, followed by an interleaved sequence of three visits each to Titan and the phase calibrator J1751-1950. The phase-center was updated in real-time to track Titan's moving position on the sky. The flux of the calibration quasar J1733-1304 was measured on 2017-05-04 and 2017-05-17, with an accuracy of $\pm6$\%, implying a similar accuracy in the flux scale of our calibrated spectra.

Data were flagged and calibrated in CASA 5.1 \citep{jae08} using the automated pipeline scripts supplied by the Joint ALMA Observatory \citep[see][]{shi15}. The data from May 8 and May 16 were combined after regridding and Doppler-shifting to Titan's rest frame. A small offset in declination (of unknown origin) was identified and corrected in Titan's position with respect to the ALMA phase center on May 8 ($0.06''$ in setting 1 and $0.11''$ in setting 2). Titan's continuum flux was subtracted using low-order polynomial fits to the (line-free) spectral regions adjacent to our lines of interest. Imaging and deconvolution were performed using the {\tt Clark clean} algorithm with a pixel size of $0.025''$, a flux threshold of twice the expected RMS noise and a mask diameter of $1.3''$ (8700~km at Titan's distance of 9.26~AU), encircling the entirety of the detected flux from Titan's atmosphere.

The coordinate scales of the cleaned images were transformed to physical distances with respect to the center of Titan. Titan's north pole was oriented $5.3^{\circ}$ counter-clockwise from celestial north, and tilted towards the observer by $26^{\circ}$. This is close to the maximum polar tilt due to the proximity of our observations to Titan's southern winter solstice on 2017-05-24.

\section{Cassini Observations}

Cassini's Composite Infra-Red Spectrometer (CIRS) is a Fourier transfer spectrometer covering 10-1500~cm$^{-1}$ over three focal planes: FP1 (10-600~cm$^{-1}$); FP3 (600-1100~cm$^{-1}$); and FP4 (1100-1500~cm$^{-1}$).
For additional details regarding the CIRS instrument see \citet{fla04} and \citet{jen17}.
Here, we use data obtained using the FP3 and FP4 focal planes, which comprise linear arrays of 10 pixels each with a size of 0.27~mrad \citep{nix09b}. These focal planes permit high spatial resolution nadir mapping of Titan's disk, as well as limb profiling of the atmosphere. The data presented here were obtained on orbits \#261 (2017-02-17) and \#275 (2017-05-24), the latter being Cassini's final flyby of Titan on which whole-hemisphere CIRS (nadir) mapping was performed.

The CIRS nadir observations were taken with moderate (2.5~cm$^{-1}$) spectral resolution, which allowed rapid mapping of an entire hemisphere of Titan by scanning the FP3/4 pixel arrays in a `pushbroom' pointing sequence. Observations were obtained on orbit \#275 at a distance from Titan of $\sim$300\,000~km, which resulted in a pixel size of $\sim$100~km, or $\sim$3$^{\circ}$ of latitude when projected to Titan's location. FP3 contained the prominent rovibrational emission bands of HCN, HC$_3$N, C$_2$H$_2$, C$_2$H$_6$, C$_3$H$_4$ and C$_4$H$_2$ in the range 628-814~cm$^{-1}$ (see \citealt{tea09} for example spectra). The interval 650-660~cm$^{-1}$ is free from trace gas emission features and was used to determine the continuum level. In the present study, we focus primarily on the HCN and HC$_3$N data for comparison with ALMA observations of these species.

CIRS limb observations were obtained on orbits \#261 and \#275 with the FP3 and FP4 pixel arrays orientated perpendicular to the limb to give spectra covering tangent altitudes 0-600~km.
Observations were in `sit-and-stare' mode at 0.5~cm$^{-1}$ spectral resolution, resulting in high signal-to-noise, high-resolution spectra at a single latitude. Orbit \#261 covered the latitude $-11^{\circ}$(S) with a field-of-view size of $\sim$45~km and \#275 covered latitude $-51^{\circ}$(S) with a field-of-view size of $\sim$60~km. On orbit 261 only every other pixel was measured, so this observation had an equivalent vertical resolution of $\sim$120~km.

%\newpage
\section{Results and Analysis}

\subsection{Spectra and Maps}
\label{sec:specmap}

\begin{figure*}
\centering
\includegraphics[width=0.49\textwidth]{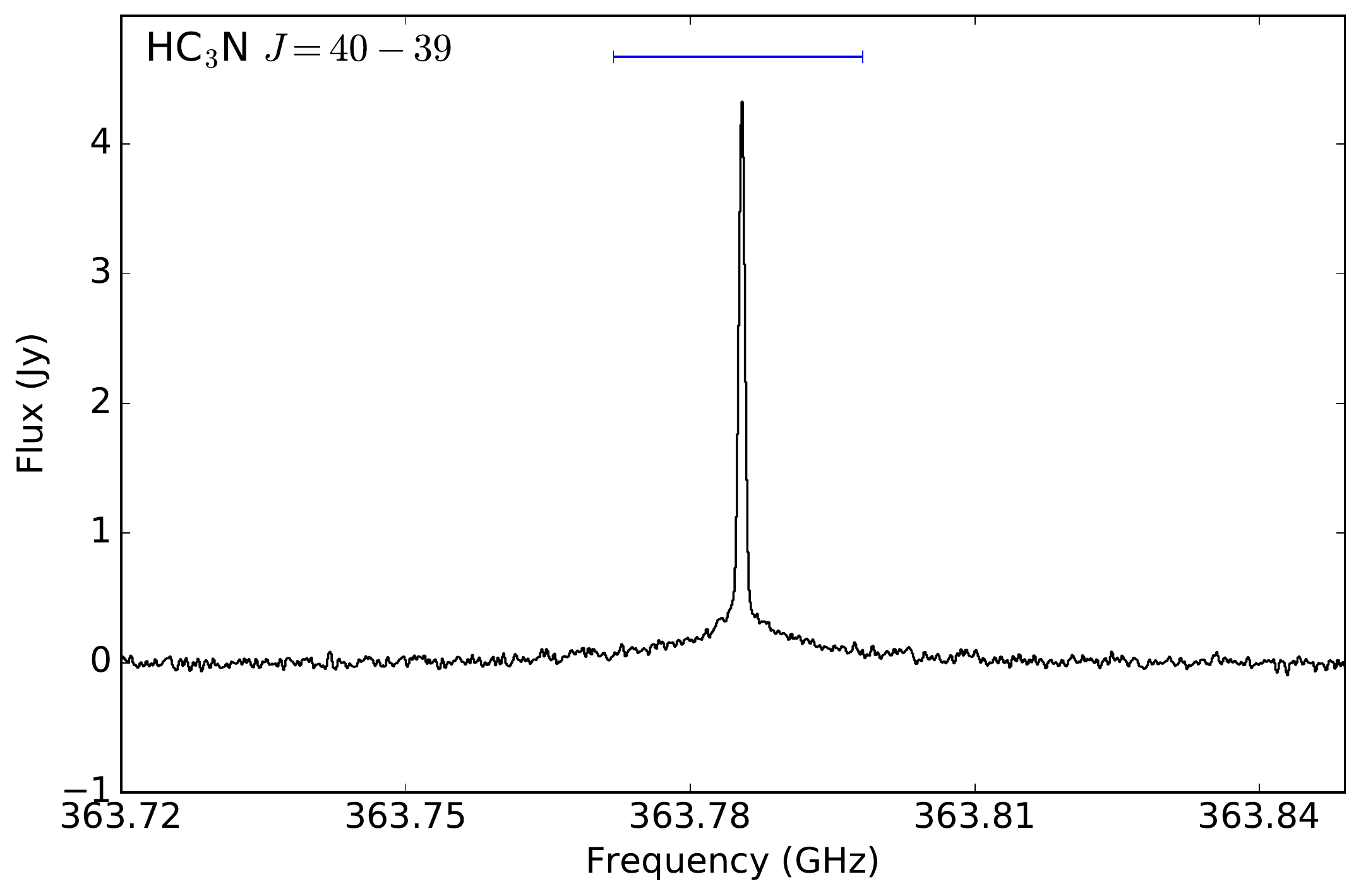}
\hspace{2mm}
\includegraphics[width=0.42\textwidth]{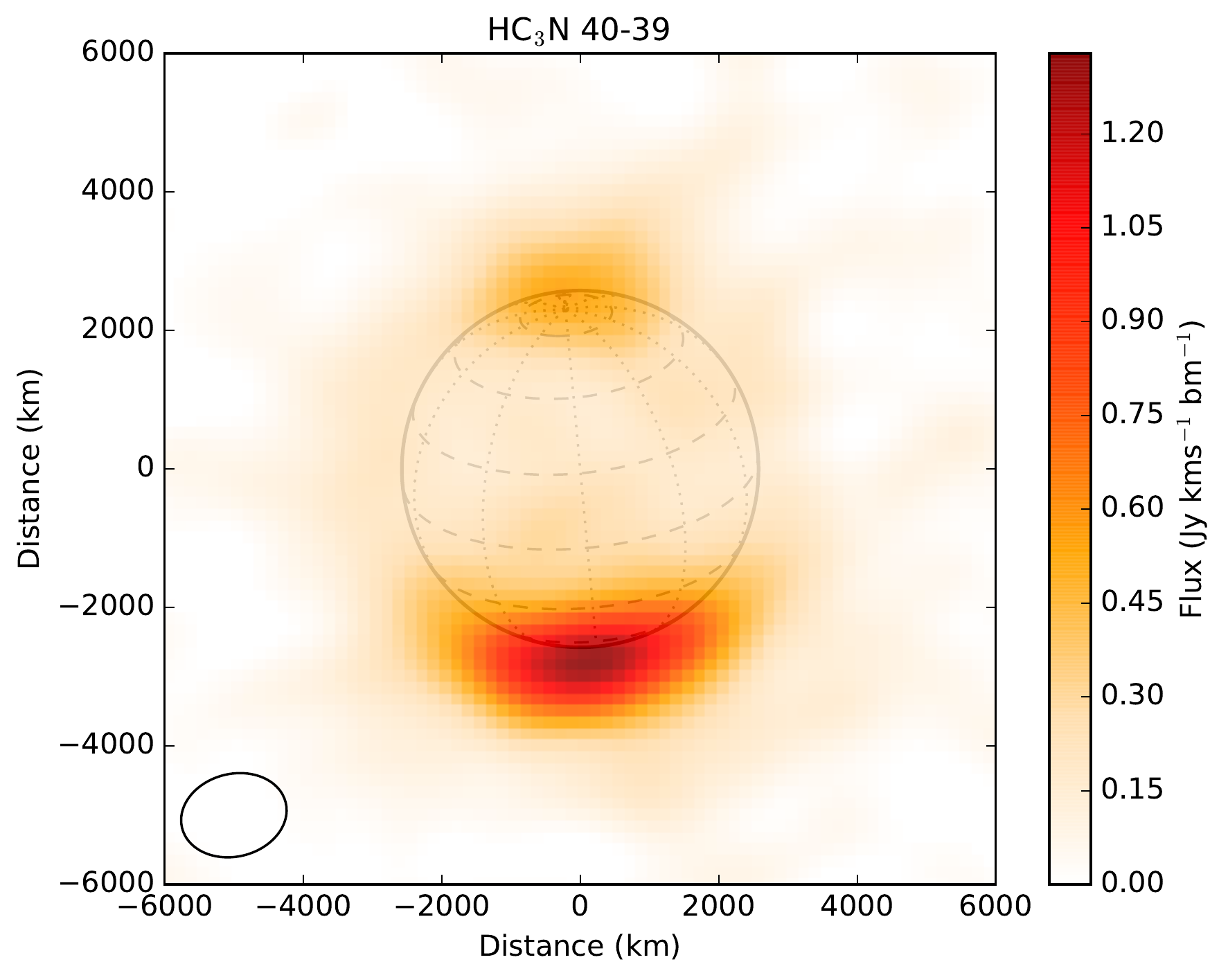}
\includegraphics[width=0.49\textwidth]{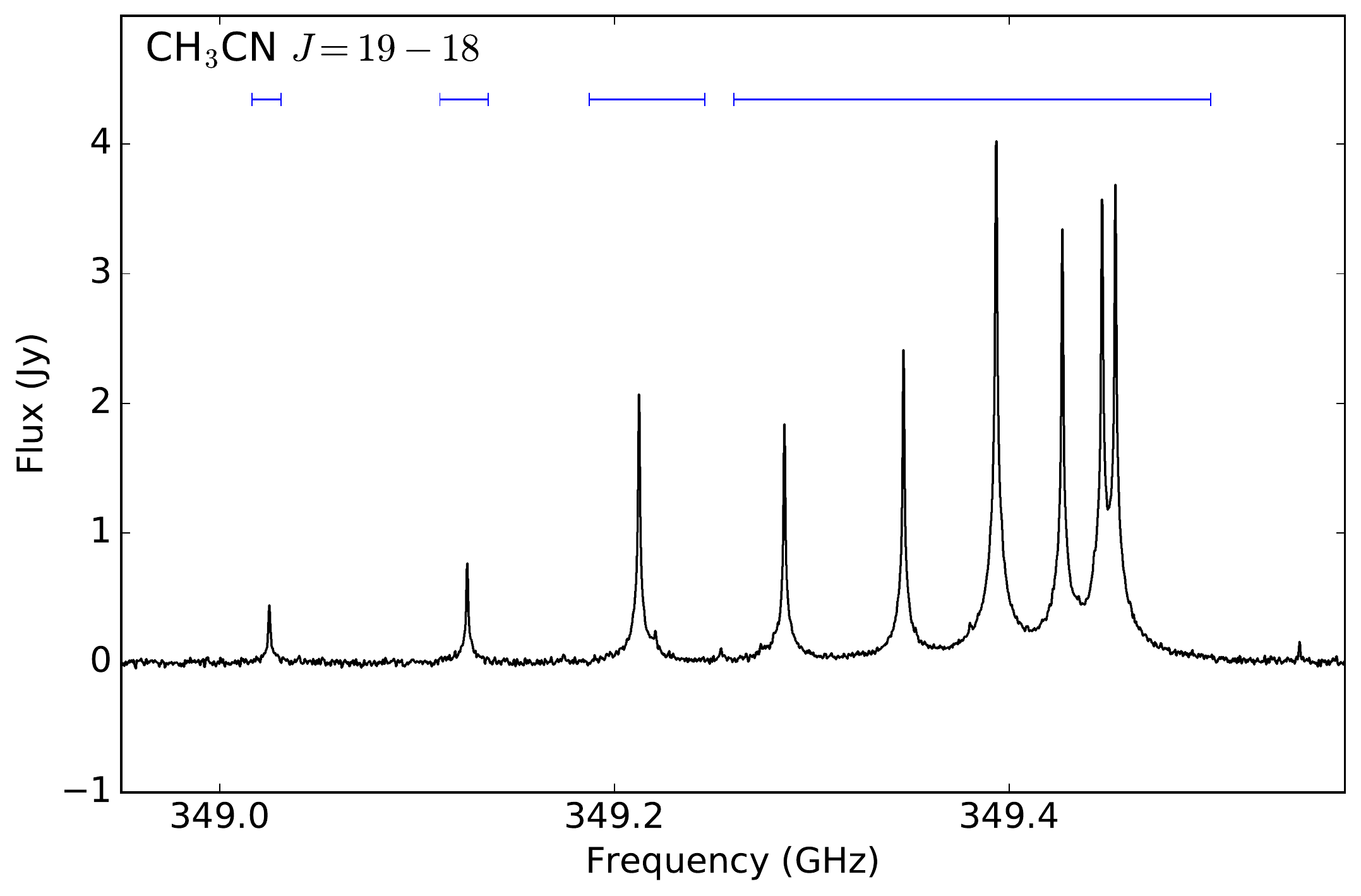}
\hspace{2mm}
\includegraphics[width=0.41\textwidth]{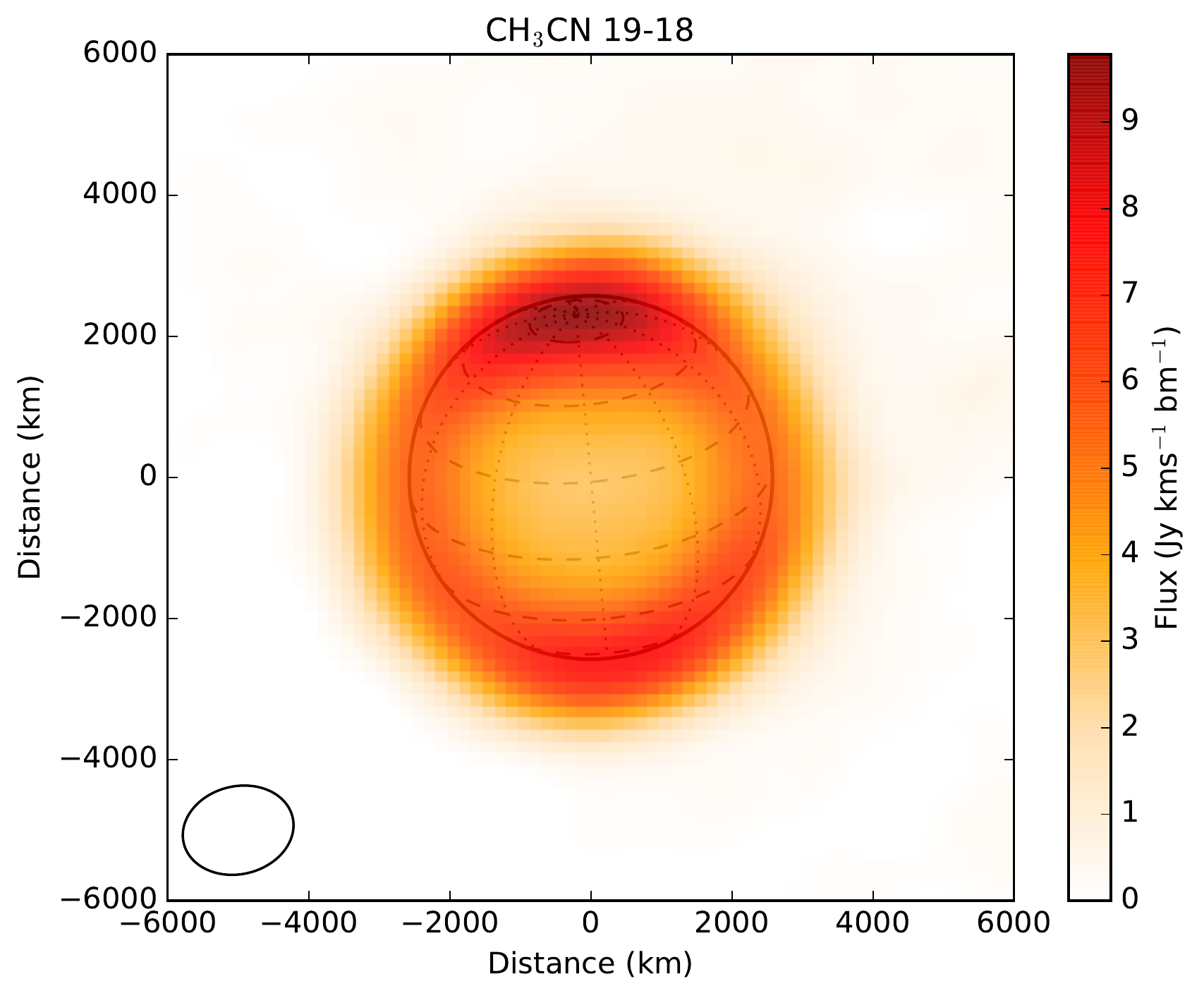}
\includegraphics[width=0.49\textwidth]{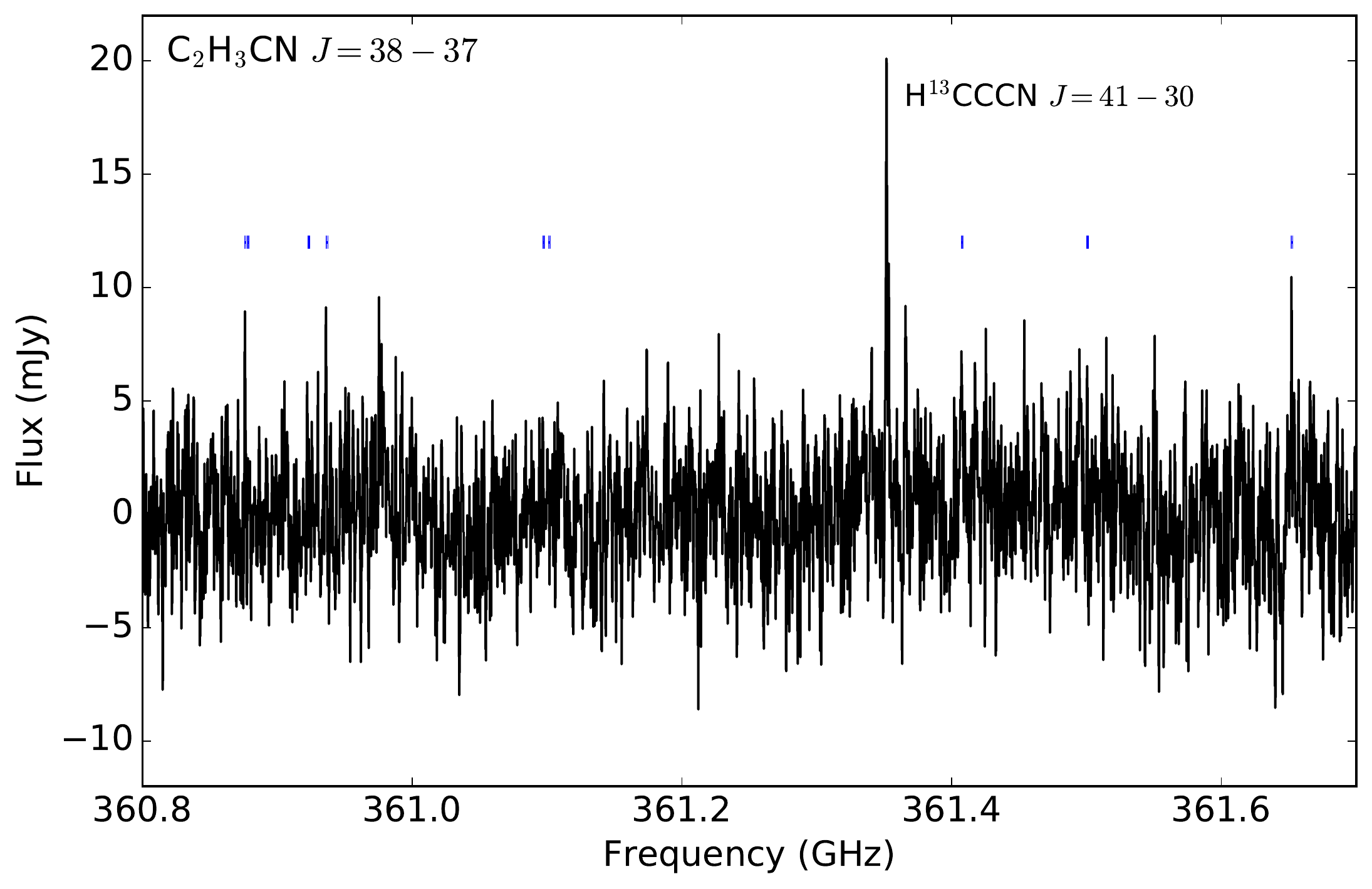}
\hspace{2mm}
\includegraphics[width=0.42\textwidth]{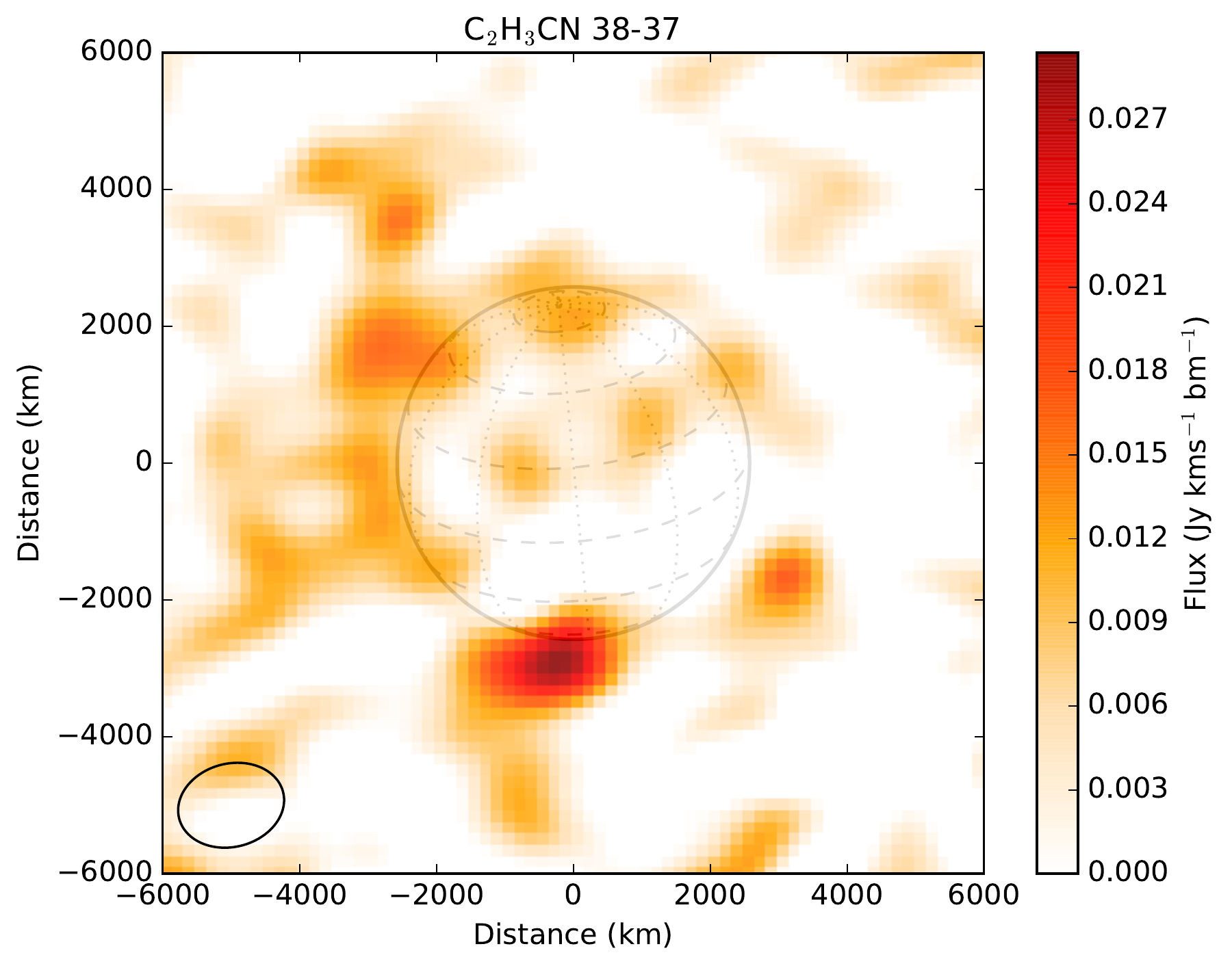}
\includegraphics[width=0.49\textwidth]{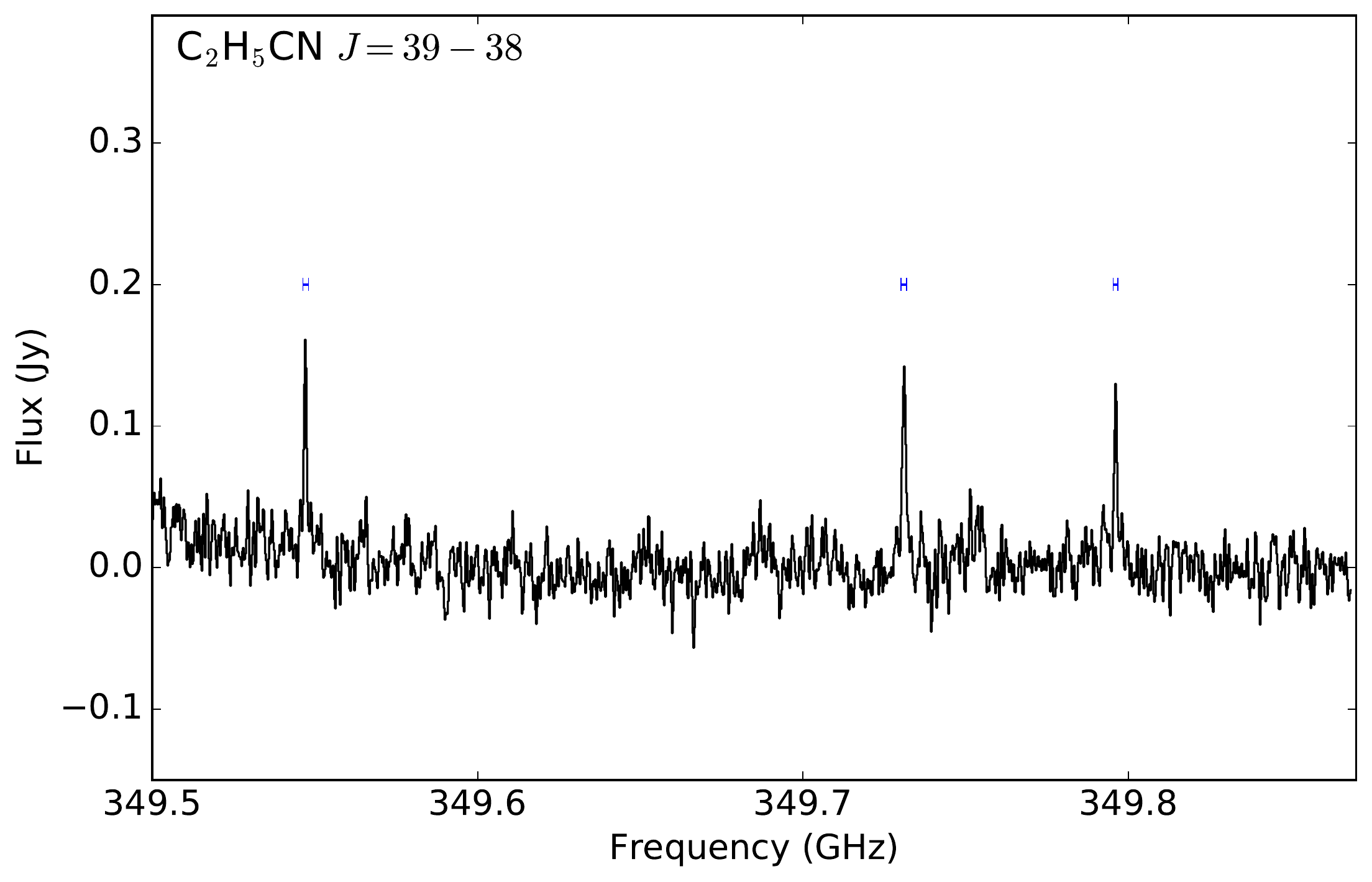}
\hspace{2mm}
\includegraphics[width=0.42\textwidth]{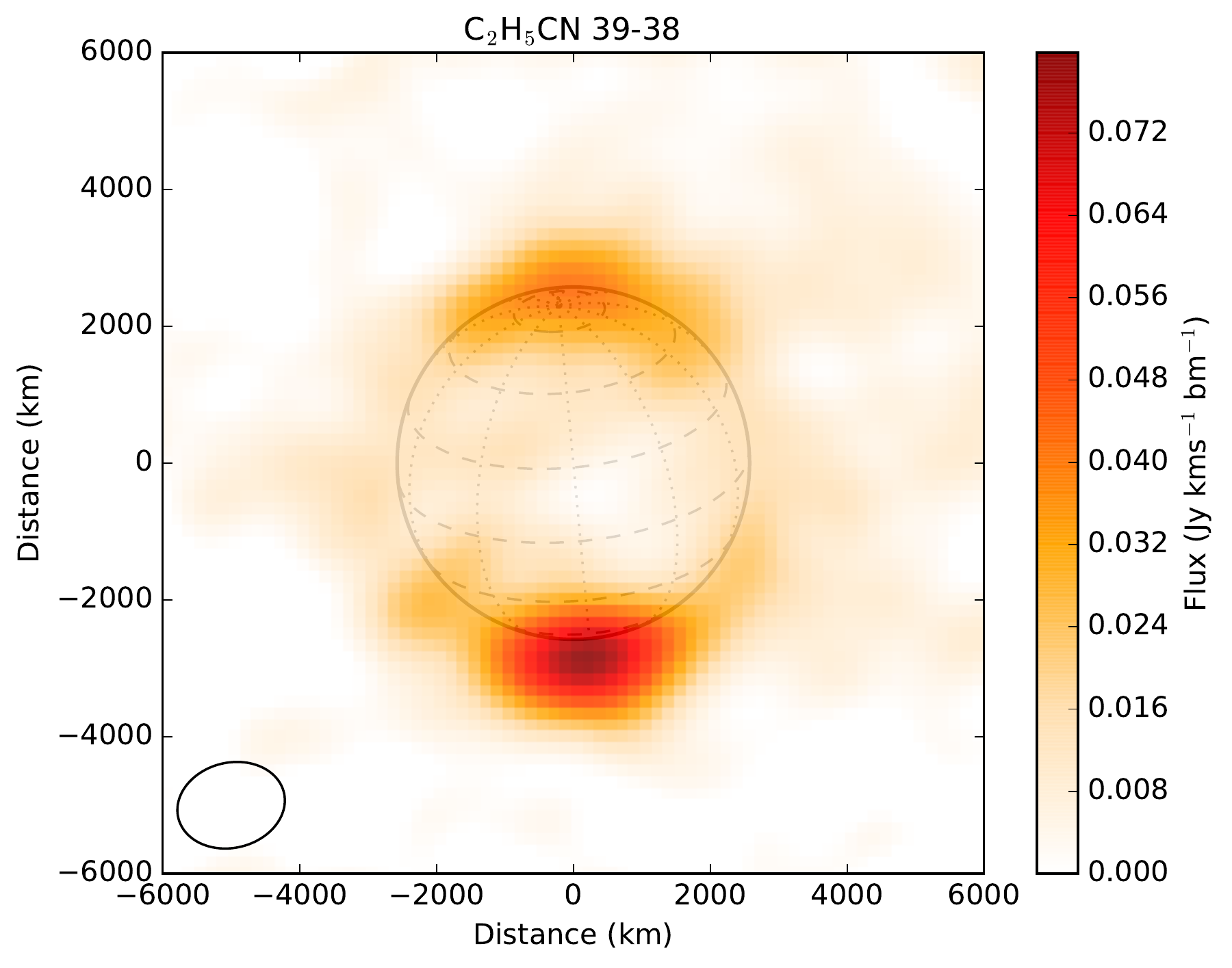}
\caption{Continuum-subtracted molecular spectra and integrated emission maps for HC$_3$N, CH$_3$CN, C$_2$H$_3$CN and C$_2$H$_5$CN, observed using ALMA in 2017 May. Spectra (apart from C$_2$H$_3$CN) were integrated over a circular aperture with radius 4300 km, centered on Titan. The C$_2$H$_3$CN spectrum was extracted from a beam centered at the S-polar emission peak for that molecule. Spectral integration range(s) for the emission maps are shown with blue bars. The grey wireframe spheres indicate Titan's surface and alignment in the field of view. Ellipse (lower-left) shows the spatial resolution (beam FWHM) for each map. \label{fig:maps1}}
\end{figure*}

\begin{figure*}
\centering
\includegraphics[width=0.49\textwidth]{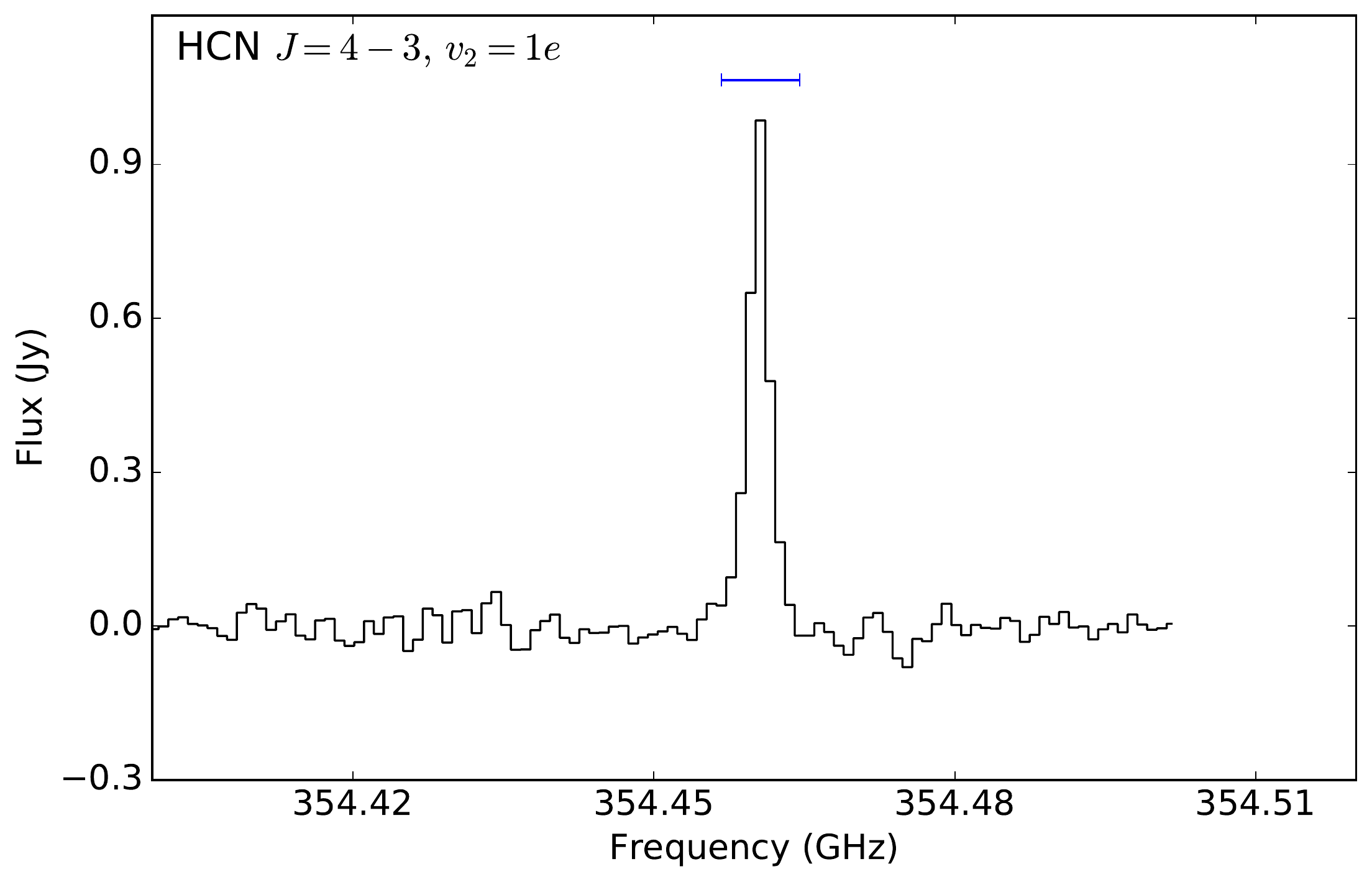}
\hspace{2mm}
\includegraphics[width=0.42\textwidth]{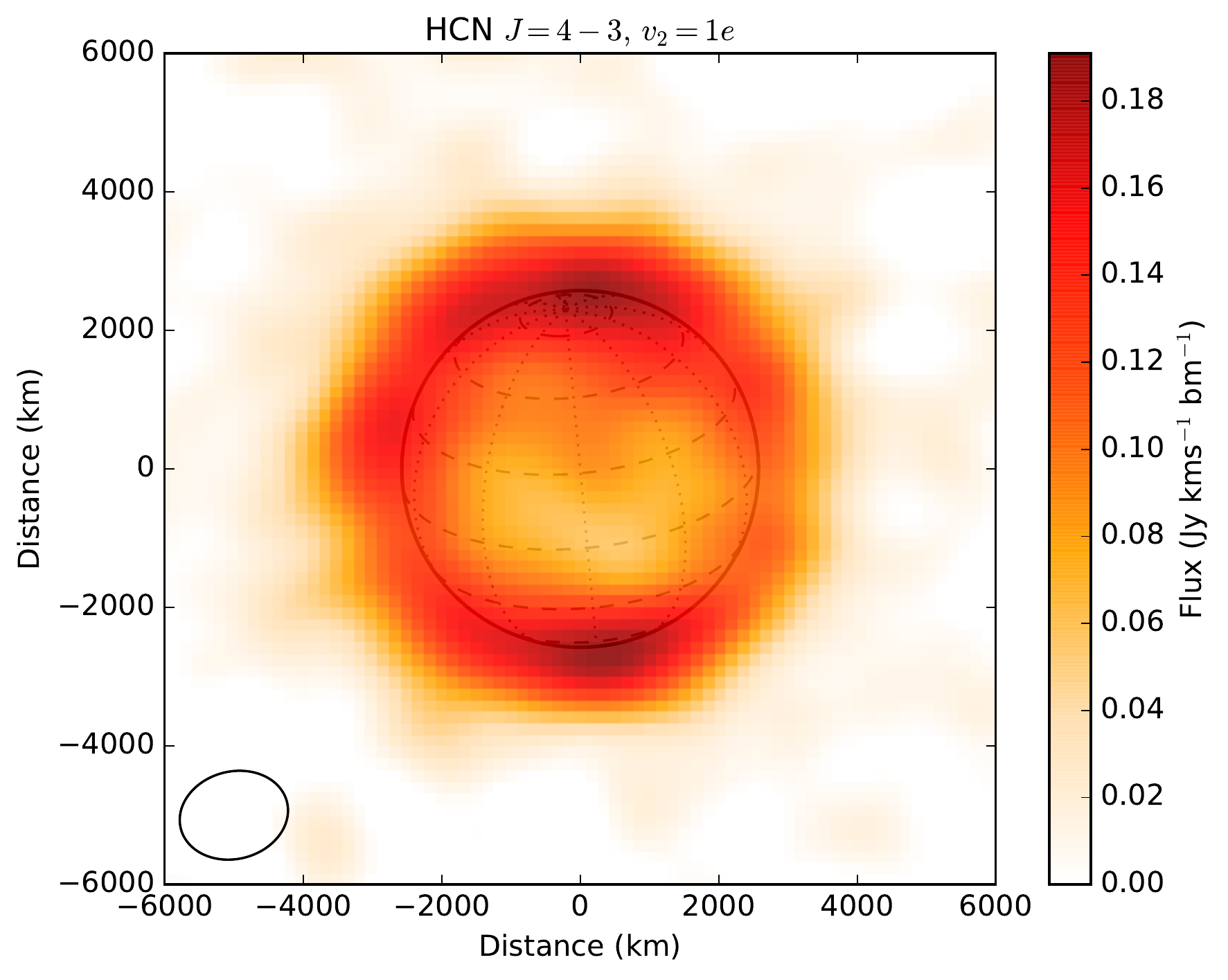}
\includegraphics[width=0.49\textwidth]{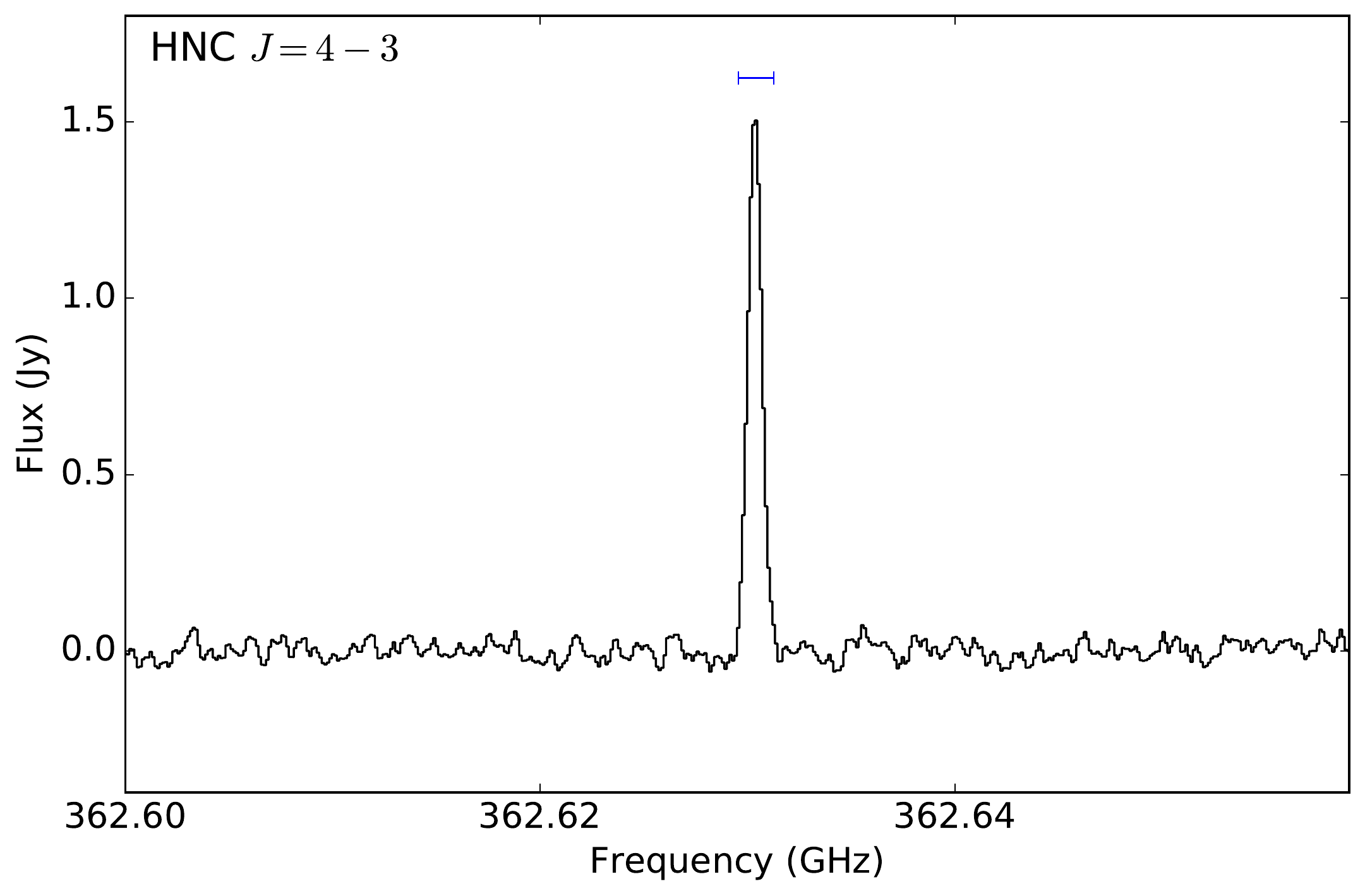}
\hspace{2mm}
\includegraphics[width=0.42\textwidth]{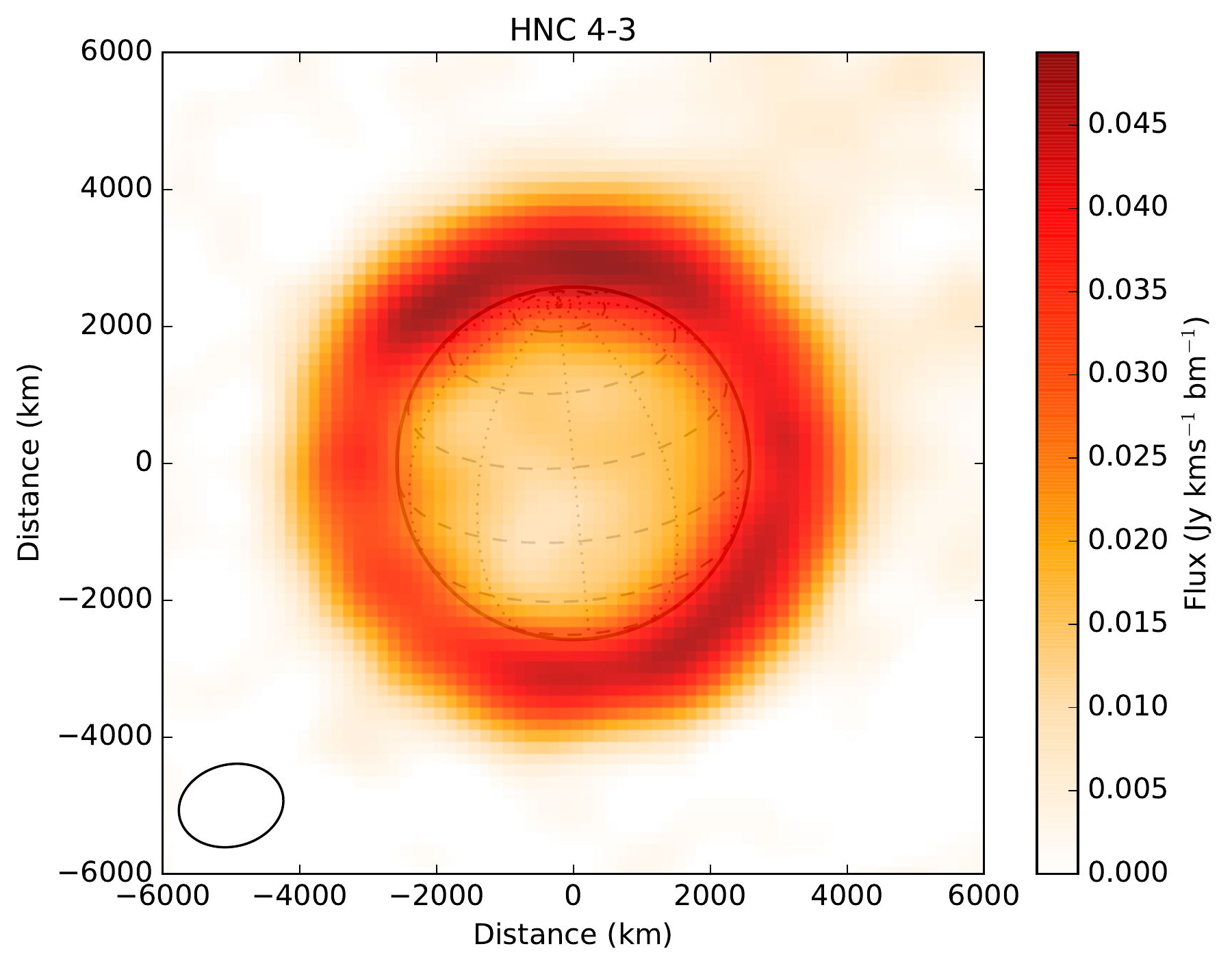}
\includegraphics[width=0.49\textwidth]{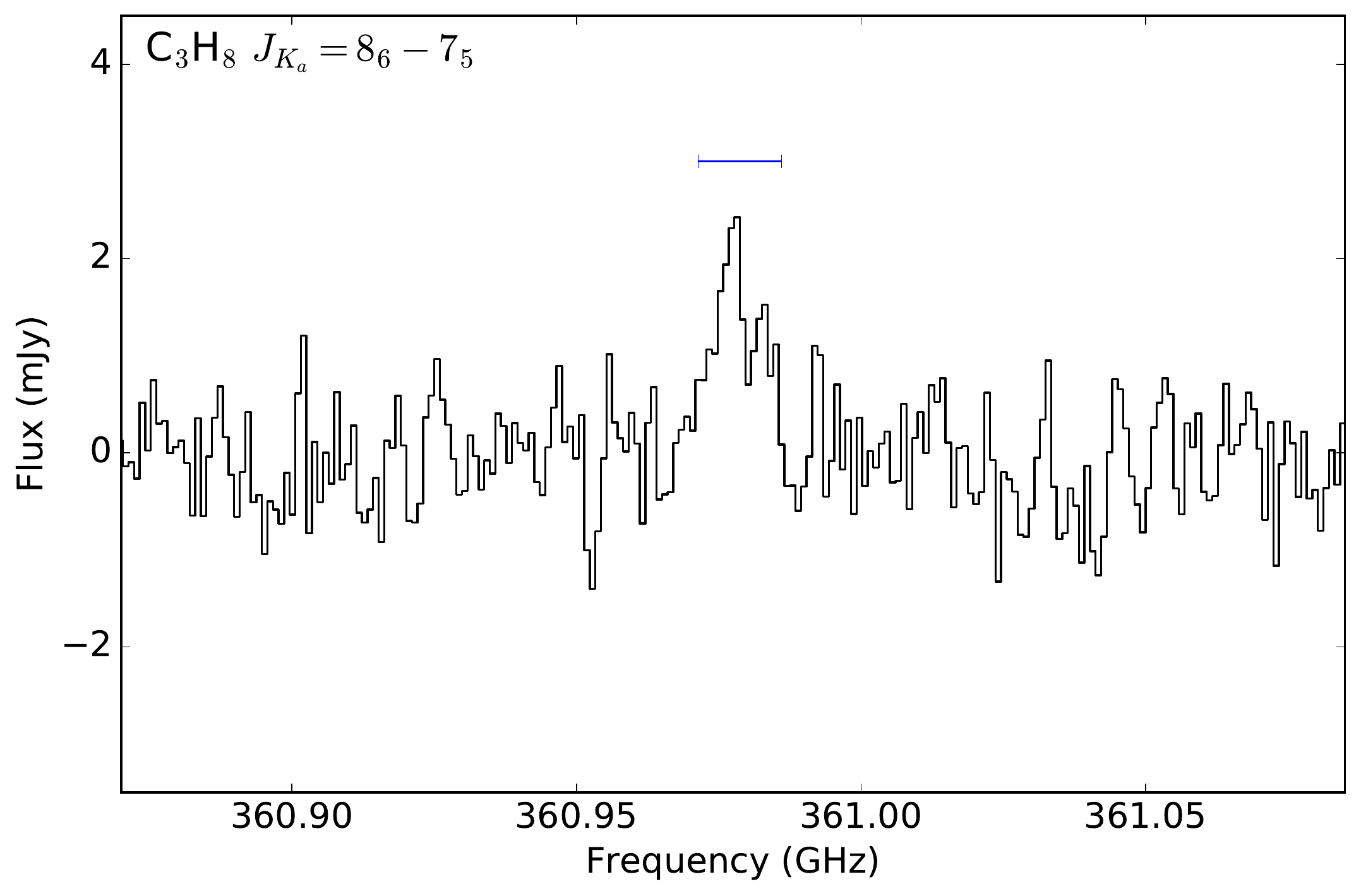}
\hspace{2mm}
\includegraphics[width=0.42\textwidth]{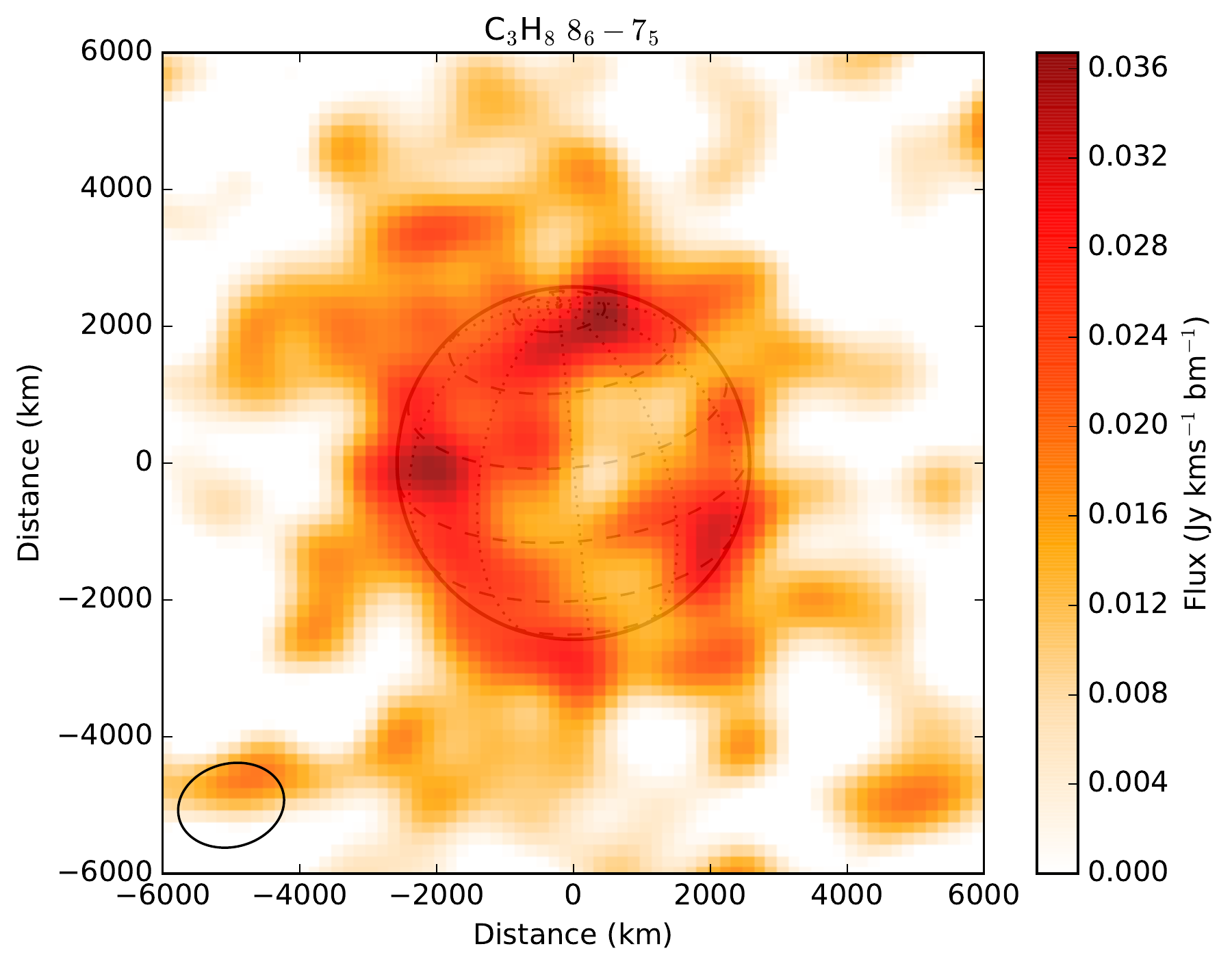}
\caption{Same as Figure \ref{fig:maps1} but for HCN, HNC and C$_3$H$_8$. \label{fig:maps2}}
\end{figure*}

ALMA spectra for our main species of interest are shown in the left panels of Figures \ref{fig:maps1} and \ref{fig:maps2}. These were obtained by integrating the reduced data cubes inside a circular aperture with radius 4300~km, centered on Titan (compared with the solid-body radius of 2575~km). This was sufficiently large to include all detected flux from the most extended (HNC) emission map.  The resulting disk-integrated spectrum was too noisy to permit detection of C$_2$H$_3$CN, so the spectrum for this species was extracted from a beam centered at the emission peak for this molecule (200~km outside Titan's southern polar limb; see Figure \ref{fig:maps1}). 

Spectral features greater than $3\sigma$ above the noise were identified and integrated to produce the emission maps shown in the right panels of Figures \ref{fig:maps1} and \ref{fig:maps2}. For \vycn, only three of the expected nine spectral lines were detected with greater than $3\sigma$ confidence (as shown in Table \ref{tab:spec}; two of these lines are blends of two closely-spaced rotational transitions).  The unexpected weakness of the \vycn\ emission could be explained by temporal variability with respect to the earlier studies of this molecule by \citet{pal17} and \citet{lai17}, but may also be due to relatively poor coupling between our ALMA beam and the intrinsic \vycn\ distribution --- the earlier studies used a larger beam size ($\sim1''$), which was more sensitive to emission spread across Titan's entire $\approx1''$ disk plus atmosphere --- whereas our $\sim0.2''$ beam was only able to recover the compact \vycn\ emission peak near the south pole. To improve the signal-to-noise (S/N) ratio of the \vycn\ map in Figure \ref{fig:maps1}, integration was performed over the frequency ranges corresponding to the nine expected strongest lines in one of our ALMA spectral windows (360.8-361.7~GHz), using the model of \citet{pal17} as a guide for the line strengths, and adopting a 1.2~\kms\ integration width for each spectral line. 

The resulting ALMA maps provide the highest-resolution (instantaneous) views of HNC, \mecn, \vycn, and \etcn\ in Titan's entire sunward-facing-hemisphere published to-date using any instrument. Characteristic emission patterns are apparent for each species, and provide for the first time, a detailed view of the global column density distributions for these gases. Since most of the detected emission lines are relatively weak (and unsaturated), the majority of the observed molecular emission is expected to be optically thin (a notable exception is \cyano, which becomes optically thick in the vicinity of the south pole; \citealt{cor18}). Titan's sub-mm continuum forms near the tropopause (at altitudes $z\approx40-60$~km; \citealt{the18}), where the temperature $T$ is close to 70~K. Our species of interest are synthesized primarily at higher altitudes (above a few hundred km; \emph{e.g.} \citealt{vui19}), where $T=160$-180~K, and are subsequently transported to the lower stratosphere where they condense out (at $z\lesssim90$~km; $T\lesssim120~K$). Across Titan's disk (and limb), the targeted spectral lines are observed in emission because these gases are much warmer than the 70~K background continuum source of Titan (and the 2.73~K cosmic microwave background). Furthermore, while Titan's atmospheric temperatures are known to vary by up to $\sim\pm20$~K with latitude \citep{ach11}, our observed lines are only weakly sensitive to temperature variations in this range \citep[see \eg][]{lai17,the19}. These ALMA emission maps therefore provide a good indication of the intrinsic column density distribution of each species.

The \cyano, \mecn, \vycn, \etcn, HCN and HNC maps obtained using ALMA all show strong variations in column density across Titan's disk. Limb brightening dominates the HCN, HNC and \mecn\ maps, characteristic of a relatively uniform latitudinal distribution, whereas the \cyano, \vycn\ and \etcn\ maps are dominated by polar emission, with the southern (winter) pole being strongest for the latter three species.

\begin{figure*}
\includegraphics[width=0.49\textwidth]{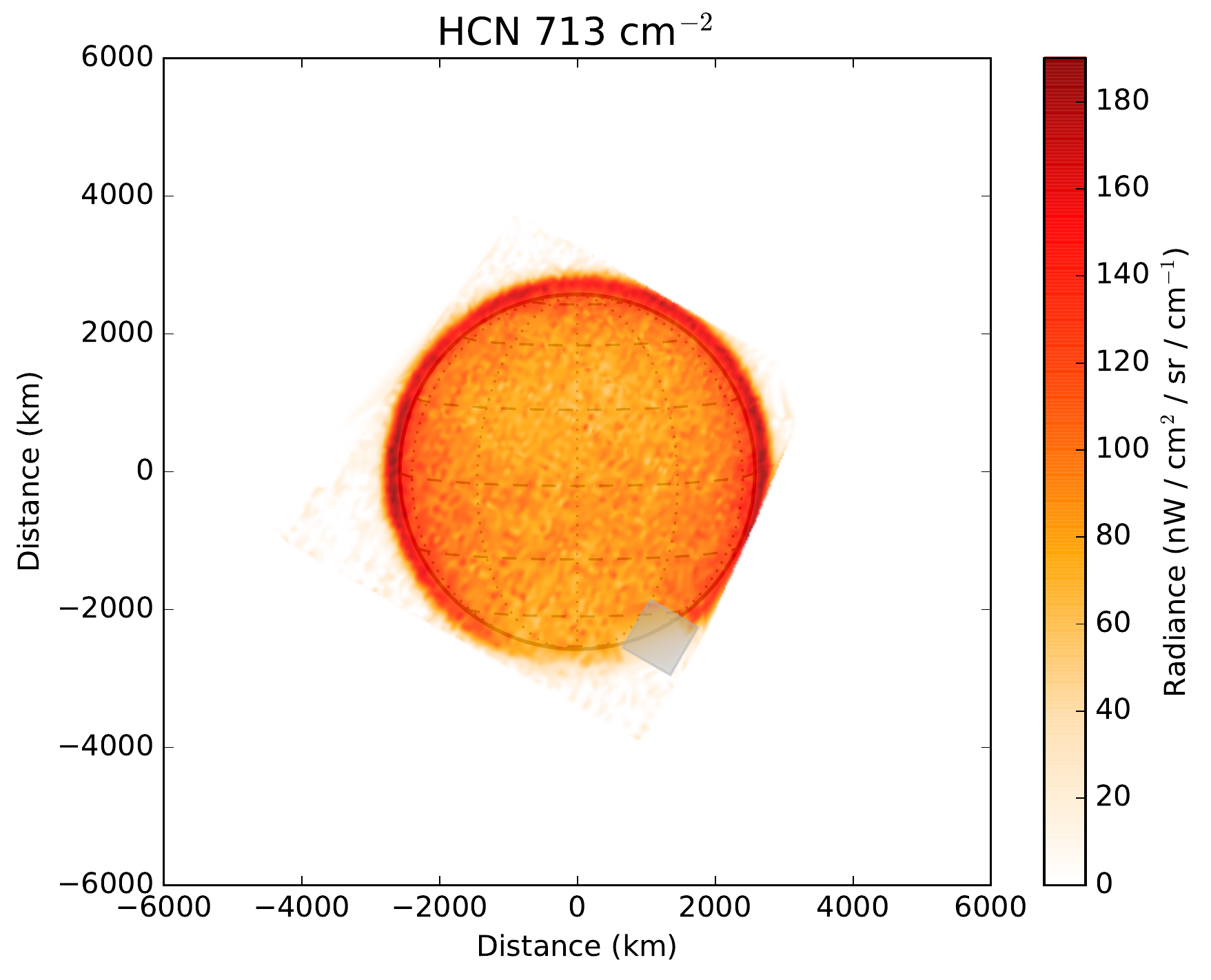}
\includegraphics[width=0.49\textwidth,trim=0 0 -3mm 0]{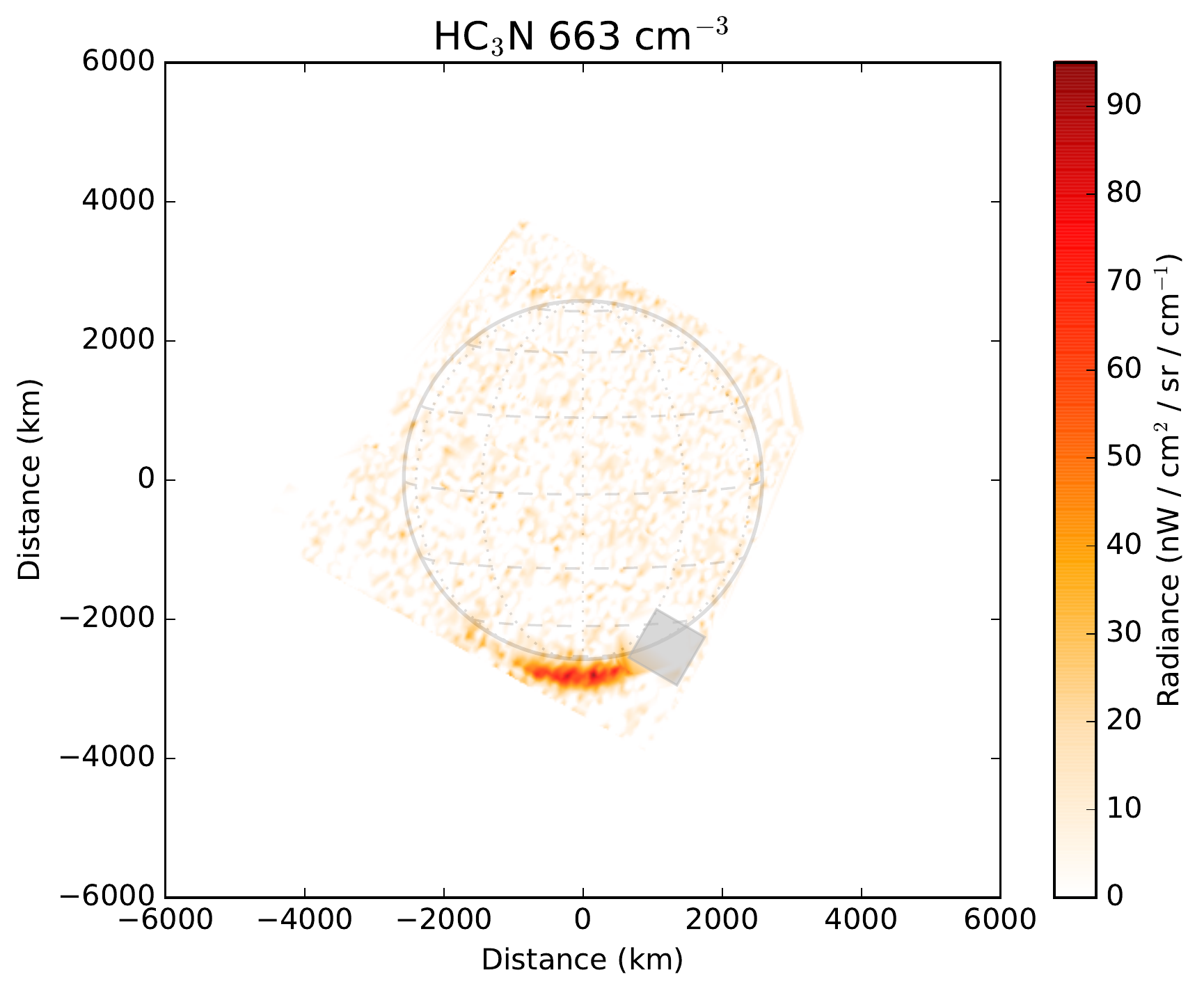}
\caption{Cassini CIRS pushbroom maps of (continuum subtracted) HCN and HC$_3$N emission from Titan, obtained during the 2017-05-24 flyby (Cassini orbit \#275). Image frequencies are given for each molecule. Grey boxes indicate missing data. \label{fig:cirsmaps}}
\end{figure*}

The HCN and HC$_3$N radiance maps obtained using Cassini CIRS (on orbit \#275) are shown in Figure \ref{fig:cirsmaps}. While the spatial resolution is significantly better than obtained by ALMA, the basic morphology of the resulting maps is similar, with a dominant south-polar emission peak for HC$_3$N (at latitudes $<-60^{\circ}$) and a more uniform, limb-brightened distribution for HCN. The intensities of these vibronic emission bands observed by CIRS have a stronger dependence on temperature than the pure rotational transitions observed by ALMA. As a result, the relationship between CIRS observed radiances and molecular column densities is non-trivial, making the interpretation of these maps more difficult. For example, it is likely that the drop in HCN radiance with decreasing latitude (south of the equator) is a consequence of falling stratospheric temperatures towards the winter pole \citep[see \eg][]{tea17,tea19}.

A unique aspect of our ALMA observations compared with typical (ground or space-based) infrared measurements is the extremely high spectral resolution (up to $\nu/\Delta{\nu}\sim10^6$), which enables the detection of emission lines from higher in the atmosphere ($z\gtrsim200$~km). This is due to the relative lack of pressure broadening above $\sim200$~km, resulting in the concentration of emission into narrow (thermally-broadened) spectral line peaks. This benefit comes with the tradeoff that we are less sensitive to low-altitude emission from the pressure-broadened line wings (below $\lesssim200$~km), although for the stronger lines (such as the pure rotational transitions of HC$_3$N and CH$_3$CN), we maintain good sensitivity down to the lower stratosphere (around $150$~km; see \eg\ \citealt{the19}). While HCN and HC$_3$N have previously been extensively mapped at higher resolution by Cassini CIRS, instantaneous, whole-hemisphere limb coverage has not been possible at high resolution using that instrument; instead, CIRS limb maps may be built up by assembling data from multiple Cassini flybys over a period of months or years (during which time the atmosphere may have changed). Furthermore, CIRS nadir mapping is most sensitive to emission from altitudes around 150~km; our ALMA maps thus provide unprecedented spatial information at higher altitudes, for the 2017 May (solstice) epoch.

While the main focus of our present study is on the nitrile observations, we also obtained a serendipitous detection of the C$_3$H$_8$ $J_{K_a}=8_6-7_5$ multiplet at 360.978~GHz (comprised of a blend of eight transitions with $\Delta K_c = 1$). This is the first time propane has been definitively detected and mapped in any extraterrestrial source at radio wavelengths (following a tentative ALMA detection on Titan by \citealt{lai17}), and is therefore worthy of further investigation. While C$_3$H$_8$ has an extremely weak radio spectrum due to its small dipole moment and large partition function, the large abundance of this molecule in Titan's atmosphere enables its detection using ALMA. Vibrational emission from C$_3$H$_8$ was previously detected on Titan by Voyager 1 \citep{mag81}, ISO \citep{cou03} and Cassini \citep{nix09}. Our ALMA map shows a relatively uniform distribution for C$_3$H$_8$ across Titan's disk, with little evidence for any latitudinal variability above the (admittedly high) noise level. 

Maps and spectra of the HC$_3$N vibrationally excited lines and isotopologues have already been presented by \citet{cor18}, and are consistent with the overall distribution of the HC$_3$N $J=40-39$ line in Figure \ref{fig:maps1}, but provide a more accurate representation of the HC$_3$N column density distribution (albeit at lower S/N), due to their reduced optical depth (see Section \ref{sec:lat}). We also obtained high-sensitivity CO and HCN spectra (including the main $J=4-3$ line of HCN as well as its $^{13}$C and $^{15}$N isotopologues). Unfortunately, these lines were too optically thick to provide useful emission maps, and require detailed radiative transfer modeling for their analysis, which will be presented in a future article. We use the vibrationally excited HCN ($J=4-3,v_2=1e$) line as an alternative to mapping the ground-state ($J=4-3$) line of this molecule, the core of which is completely optically thick (self-absorbed) across much of Titan's disk. The HCN $v_2=1e$ line is predominantly optically thin, but sits atop the strong, pressure-broadened wing of the $J=4-3$ (ground state) HCN line wing, which was subtracted using a polynomial fit, thus allowing the $v_2=1e$ line to be mapped.

\subsection{Latitudinal Profiles}
\label{sec:lat}

\begin{figure*}
\centering
\includegraphics[width=0.32\textwidth]{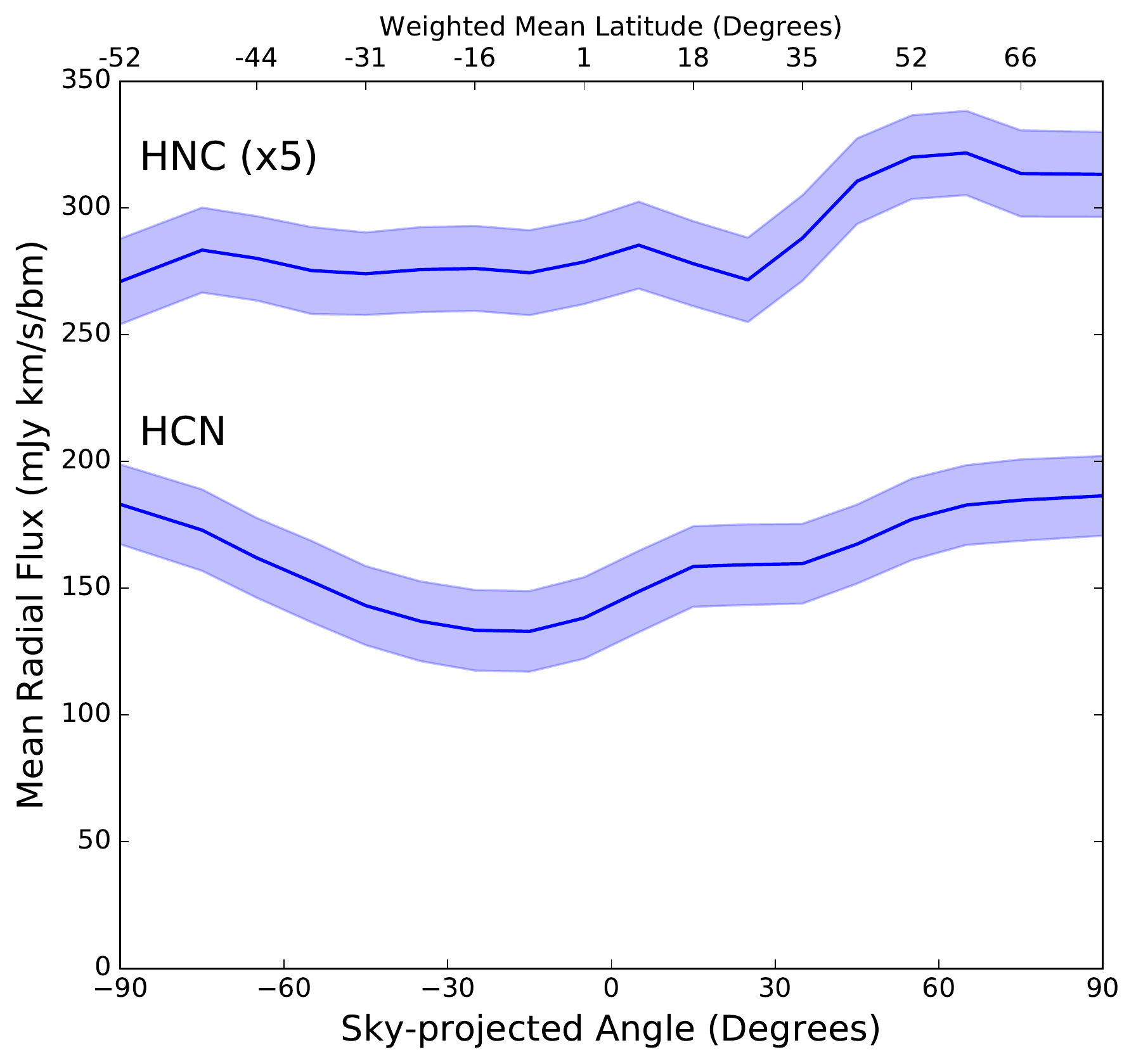}
\includegraphics[width=0.32\textwidth]{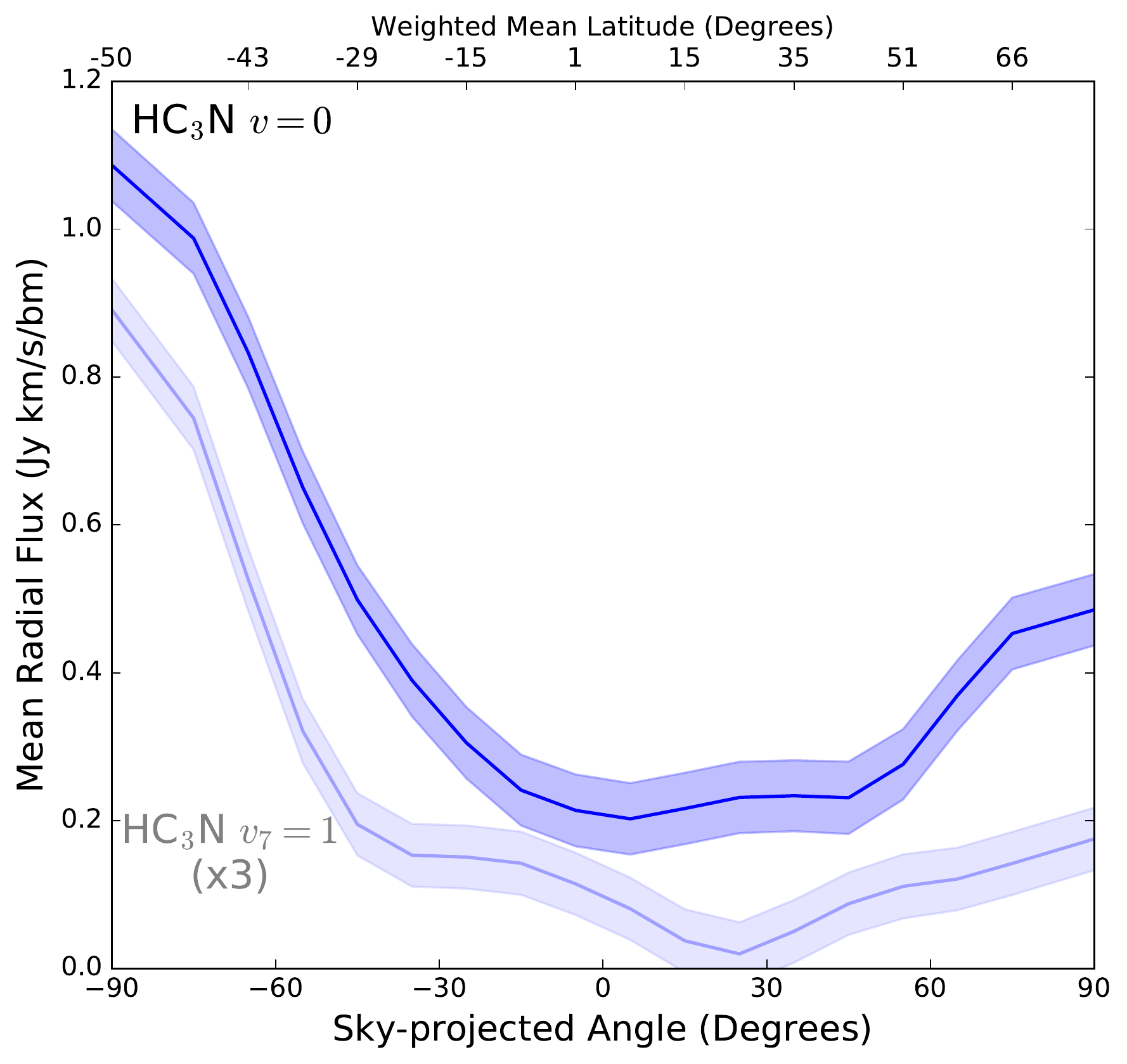}
\includegraphics[width=0.32\textwidth]{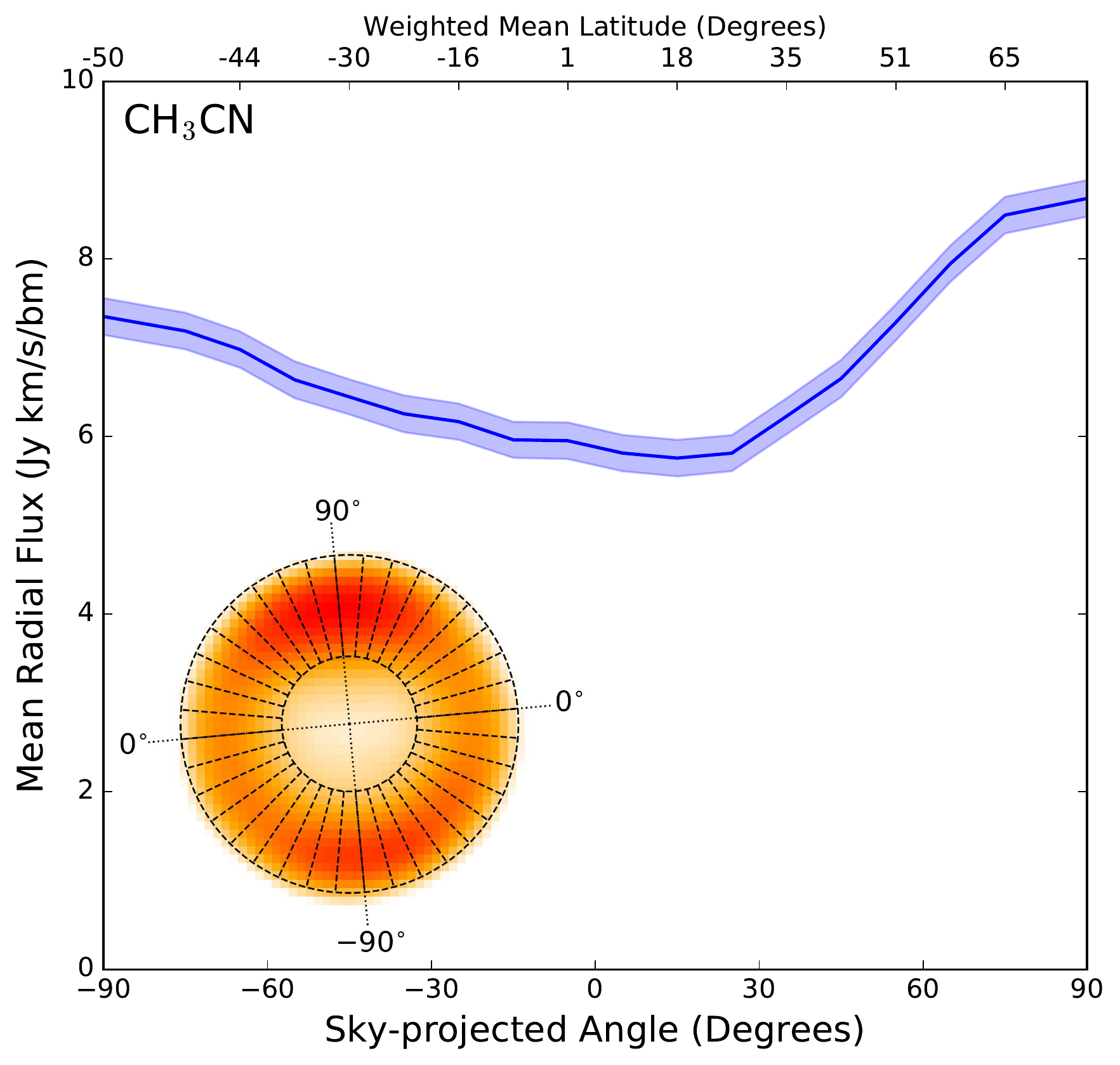}
\includegraphics[width=0.32\textwidth]{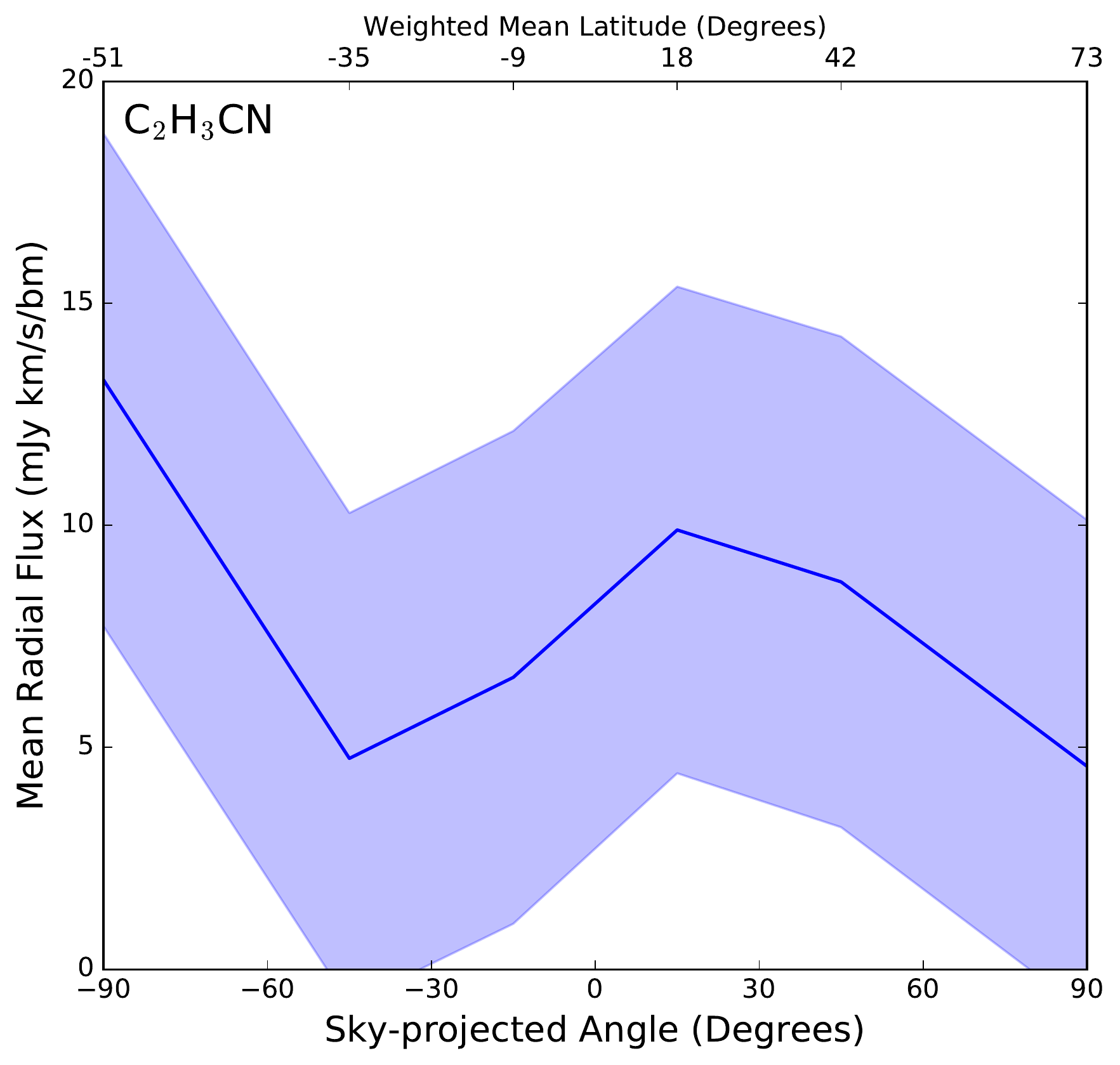}
\includegraphics[width=0.32\textwidth]{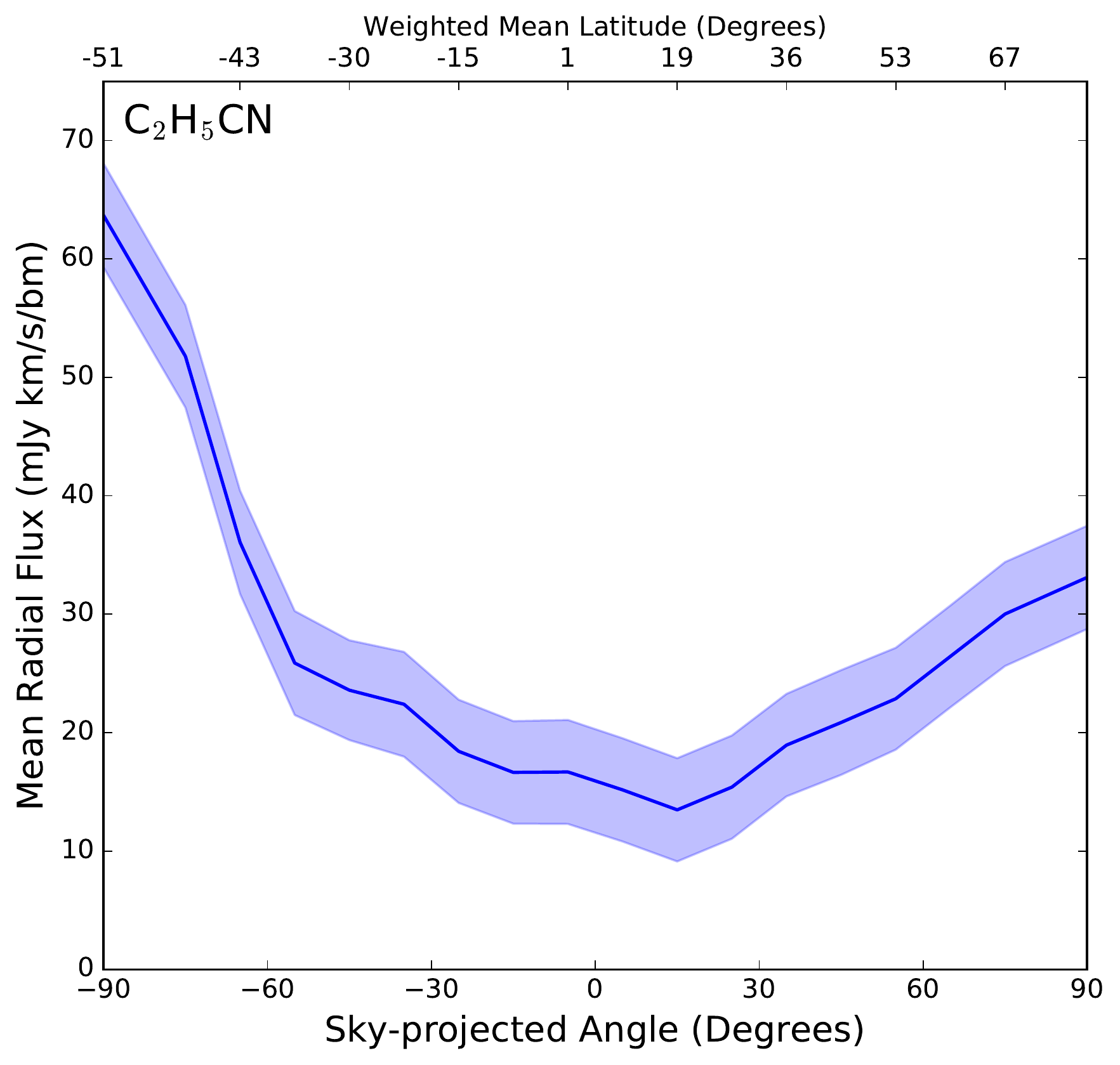}
\includegraphics[width=0.32\textwidth]{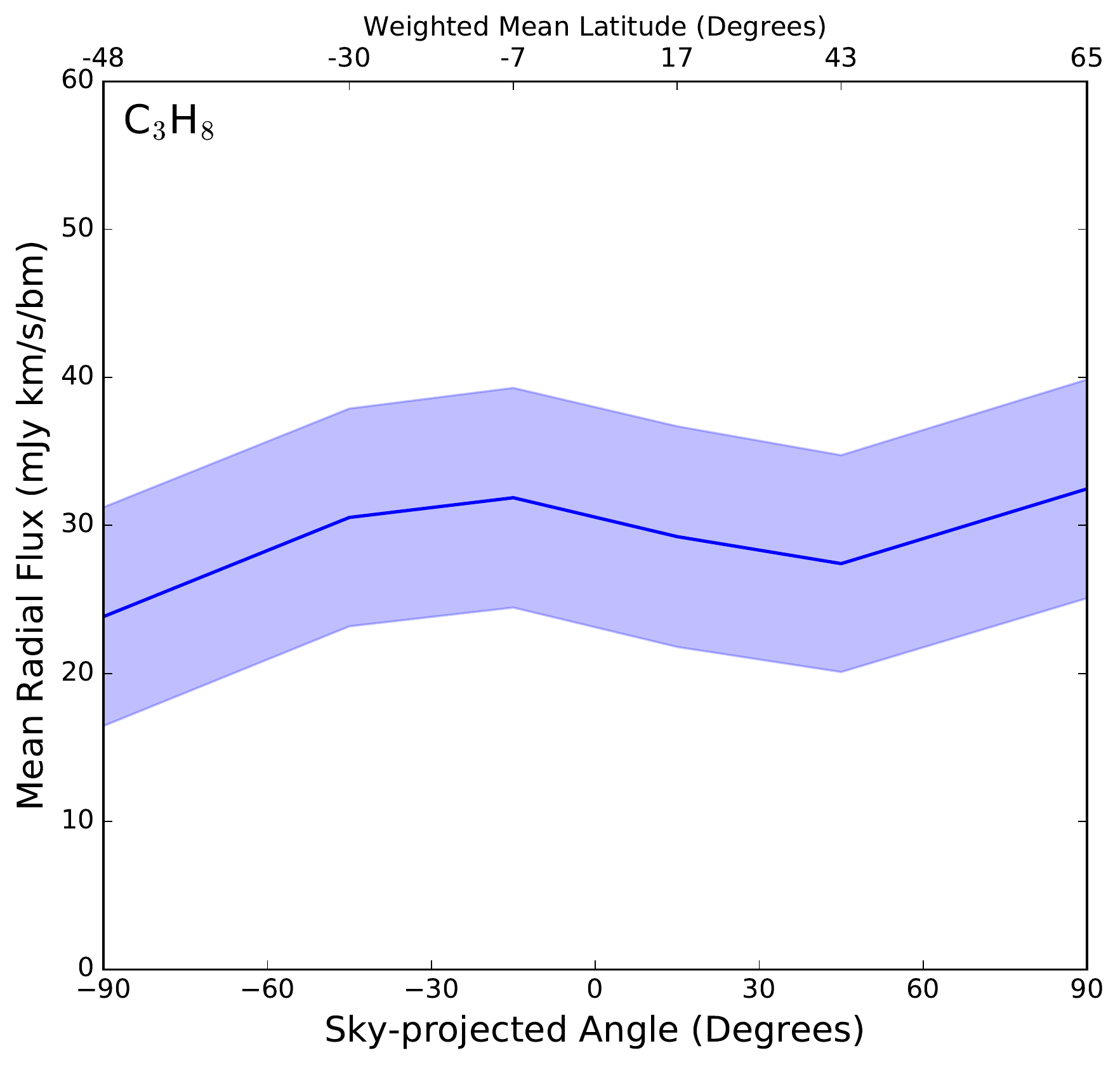}
\caption{ALMA fluxes (radially averaged) as a function of angle from the bisector of Titan's sky-projected polar axis --- the coordinate system and angular binning scheme are indicated by the inset in the upper-right panel.  Blue envelopes show the $\pm1\sigma$ error range. The weighted mean latitudes for each measurement are shown on the upper abscissa axes. HCN and HNC are shown in the same panel for comparison (with the HNC profile multiplied by 5 for display); similarly, the HC$_3$N $v_7=1$ profile is shown in a lighter colour (multiplied by a factor of 3), for comparison with HC$_3$N $v=0$. \label{fig:latprofs}}
\end{figure*}

The latitudinal variability observed in Figures \ref{fig:maps1} and \ref{fig:maps2} can be quantified by plotting the flux as a function of (sky-projected) angle about Titan's disk-center. To eliminate the impact of the differing (elliptical) beam dimensions as a function of angle, it was first necessary to degrade the resolution of our images so as to produce a circular PSF (with diameter equal to that of the long-axis of the original beam ellipse). The average flux was then obtained for each species inside radial wedges (of angular size $\phi_w$), originating from the center of Titan's disk, and plotted as a function of angle from the {perpendicular bisector of Titan's polar axis in Figure \ref{fig:latprofs}. A wedge size $\phi_w=10^{\circ}$ was used for all molecules apart from C$_2$H$_3$CN and C$_3$H$_8$, for which $30^{\circ}$ was used, to improve the S/N for these species.} The angular coordinate system and binning scheme are shown in the top-right panel for CH$_3$CN.  The relationship between the angular range covered by each wedge and the corresponding latitude on Titan is complicated due to the finite wedge size and the tilt of Titan's pole. This results in complete obscuration of the south polar region, while at the north pole, a relatively large range of latitudes are included in each wedge. The weighted mean latitude was taken within each wedge, using tangent latitudes for pixels falling outside Titan's limb. These weighted mean latitudes are shown on the upper abscissa axes in Figure \ref{fig:latprofs}.

The isomeric partners HNC and HCN are shown on the same plot (with HNC multiplied by a factor of 5 for display). The profile for vibrationally-excited HC$_3$N ($v_7=1$) is shown in addition to the ground-state ($v=0$) transition for that molecule (multiplied by 3 for display). \citet{cor18} determined that the ground-state HC$_3$N rotational lines in ALMA Band 7 become highly saturated at the south pole, which makes the $J=40-39,\ v=0$ transition unreliable as a measure of the HC$_3$N column density at that location. The $v_7=1$ transition, by contrast, is reasonably optically thin and thus provides an improved measure of the true HC$_3$N distribution.

Significant latitudinal variations are apparent for HCN, \cyano, \mecn\ and \etcn. The strongest variations are for \cyano\ and \etcn, which peak in the vicinity of the south pole and fall sharply towards the equator (over a distance smaller than the spatial resolution element). This behaviour is consistent with high concentrations of these gases within regions $\lesssim500$~km from the pole. \cyano\ and \etcn\ show a weaker peak over the north pole. By contrast, \mecn\ peaks most strongly in the north, whereas the HCN northern and southern abundance peaks are of similar strength. C$_3$H$_8$ shows no evidence for significant latitudinal variation.

\subsection{Longitudinal Variability of HNC}
\label{sec:hnc}

\begin{figure}
\centering
\includegraphics[width=\columnwidth]{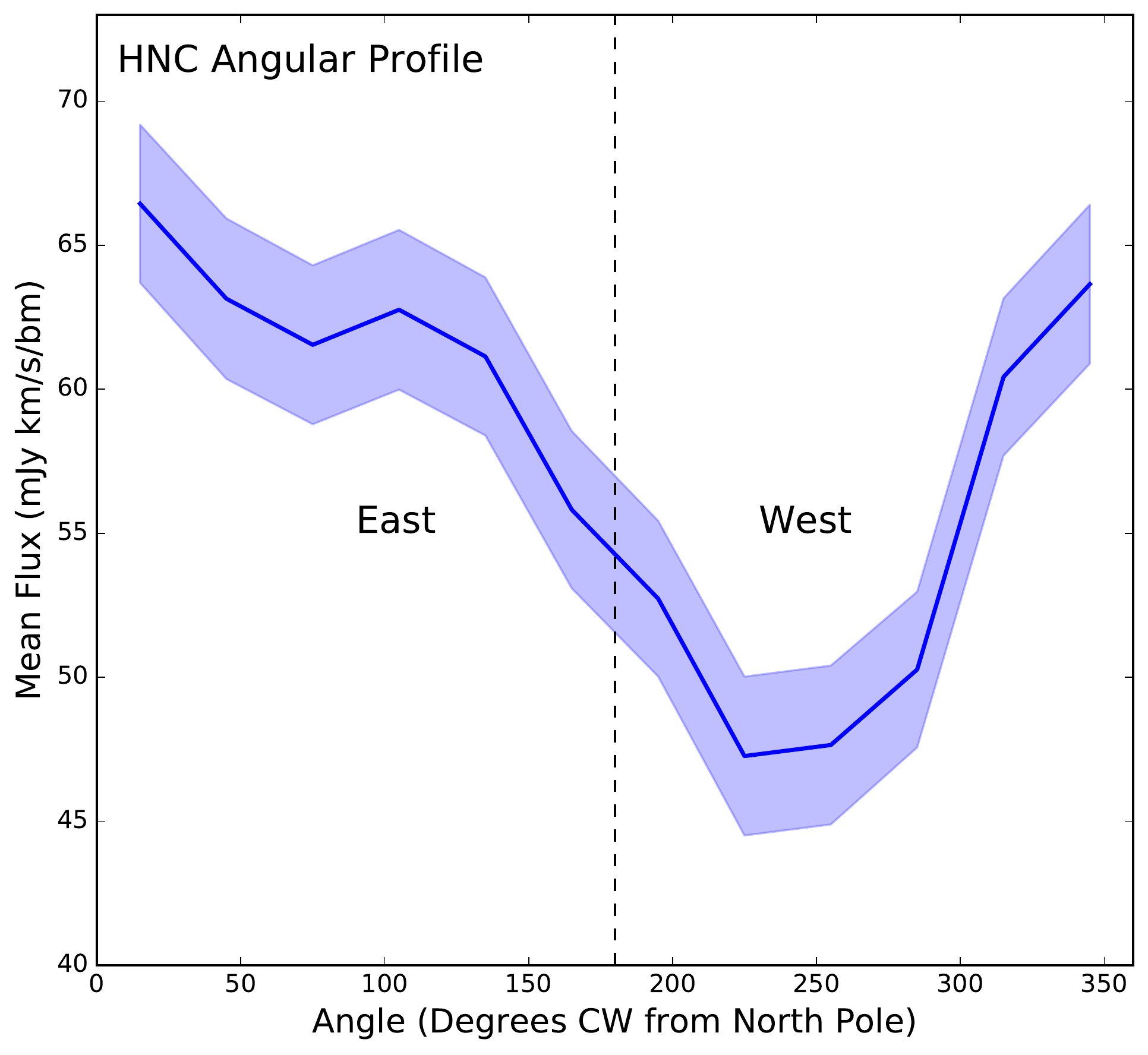}
\includegraphics[width=\columnwidth]{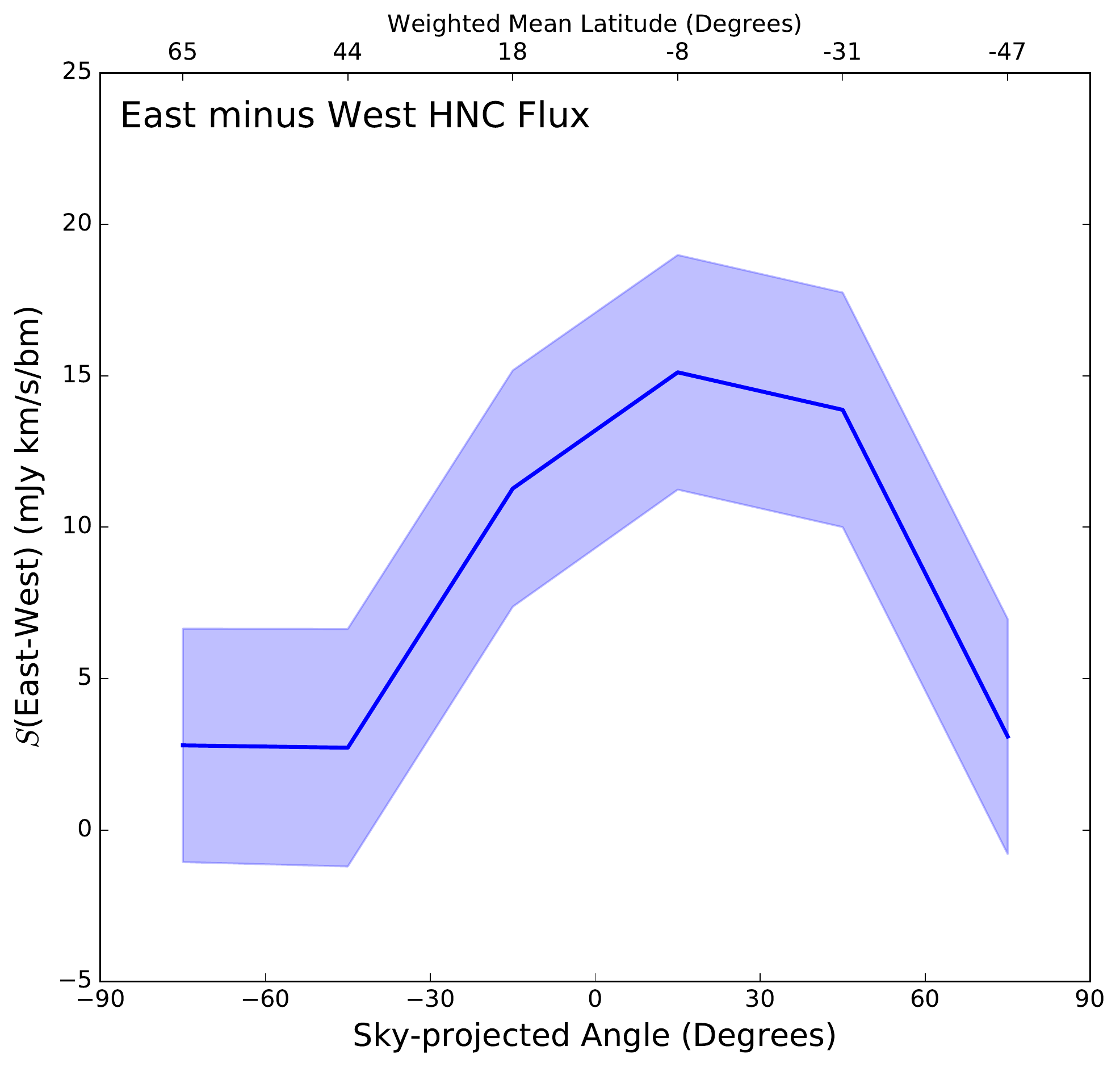}
\caption{Top: Average radial HNC flux as a function of angle from Titan's sky-projected north polar axis. Dashed line indicates 180$^{\circ}$ (\emph{i.e.} the south pole). {Bottom: Flux difference (east minus west) for HNC, as a function of angle from the bisector of Titan's sky-projected polar axis, demonstrating a clear east-west asymmetry. \label{fig:hncangular}}}
\end{figure}

The only species to show significant \emph{longitudinal} (east-west) variability is HNC, manifested by a relative weakness in the HNC emission on the western (dawn) limb. Figure \ref{fig:hncangular} (top panel) shows the average flux (in $30^{\circ}$ radial wedges), as a function of angle ($\phi$) about the disk-center, measured from the sky-projected north polar axis. The HNC emission is strongest at the north pole, relatively uniform in the eastern hemisphere, then begins to fall towards the south pole (from around $\phi=150^{\circ}$), reaching a trough in the west ($\phi=250^{\circ}$), before rising back to the northern peak. {The longitudinal asymmetry of HNC is demonstrated more clearly in the bottom panel of Figure \ref{fig:hncangular}, where the east-west flux difference ($S$(East-West)) is plotted as a function of angle from the bisector of Titan's sky-projected polar axis. Corresponding weighted-mean latitudes (shown on the upper abscissa), were calculated as in Section \ref{sec:lat}. The east-west flux difference is seen to reach a peak close to (or slightly south of) the equator. The southerly bias could be a result of viewing geometry, due to the tilt of Titan's polar axis.} 

Such longitudinal asymmetry is surprising given the presence of fast, superrotating zonal winds \citep{lel19}, which traverse Titan's globe in a period of less than a few Earth days, and are therefore expected to quickly even out atmospheric inhomogeneities as a function of longitude. While \citet{cor14} claimed the first evidence for longitudinal variations in Titan's HNC, their measurements were from relatively low resolution, low-sensitivity flux calibration data, and the observed east-west asymmetries could plausibly have resulted from the combined effects of noise and smearing by a strongly elliptical telescope beam (M. Gurwell, private communication 2016). The HNC asymmetries presented here are clearer and more robust than those identified by \citet{cor14}. We measure a total HNC flux of $S_E=383\pm7$~mJy\,\kms\ from Titan's eastern hemisphere and $S_W=338\pm7$~mJy\,\kms\ from the western hemisphere, which amounts to a ratio $S_E/S_W=(13\pm3)$\%. Thus, a conclusive, $4.3\sigma$ difference is identified in the HNC flux between the hemispheres. Use of a circularized beam renders this result free from any geometrical artifacts related to (elliptical) beam smearing.

\subsection{ALMA Limb Abundance Retrievals}
\label{sec:abund}

The variation in integrated line flux ($S$) as a function of sky-projected radial distance from Titan's limb ($r$) contains information on how the gas density ($x$) varies as a function of altitude ($z$). The vertical abundance profile $x(z)$ can thus be retrieved for each gas by fitting an atmospheric model to the ALMA observations. Despite coarse radial sampling due to limited angular resolution, our $S(r)$ measurements can nevertheless provide unique information on the high-altitude abundances for our species of interest. To generate $S(r)$ at the highest resolution (at the expense of some sensitivity), Briggs weighting of the interferometric data was used (with a $robust$ parameter of zero), leading to a resolution of $0.15\times0.13''$ ($\approx900$~km) at 362~GHz.

We generated $S(r)$ profiles by remapping the image pixels onto a polar grid with origin at the center of Titan's disk. Fluxes as a function of radius were extracted and averaged within angular regions $\phi=45$-135$^{\circ}$ and $\phi=225$-315$^{\circ}$. These ranges were chosen in order to capture a sufficiently large range of equatorial latitudes for maximum S/N while excluding the more complex polar regions (seen for \cyano, \mecn, \vycn\ and \etcn\ in Figure \ref{fig:maps1}). Abundance enhancements at the winter pole are associated with subsidence within a highly localized region inside Titan's polar vortex \citep{tea08,vin15,tea17}. This leads to downwelling of photochemically-enriched gases, resulting in a distinctly different abundance (and temperature) profile at the winter pole compared with equatorial regions. Our simplistic, 1D modeling procedure provides an estimate of the mean abundance profile at equatorial latitudes (where the abundances are dominated by high-altitude chemistry), to the exclusion of the polar regions, which are strongly influenced by seasonally-variable gas dynamics.

\begin{figure}
\centering
\includegraphics[width=\columnwidth]{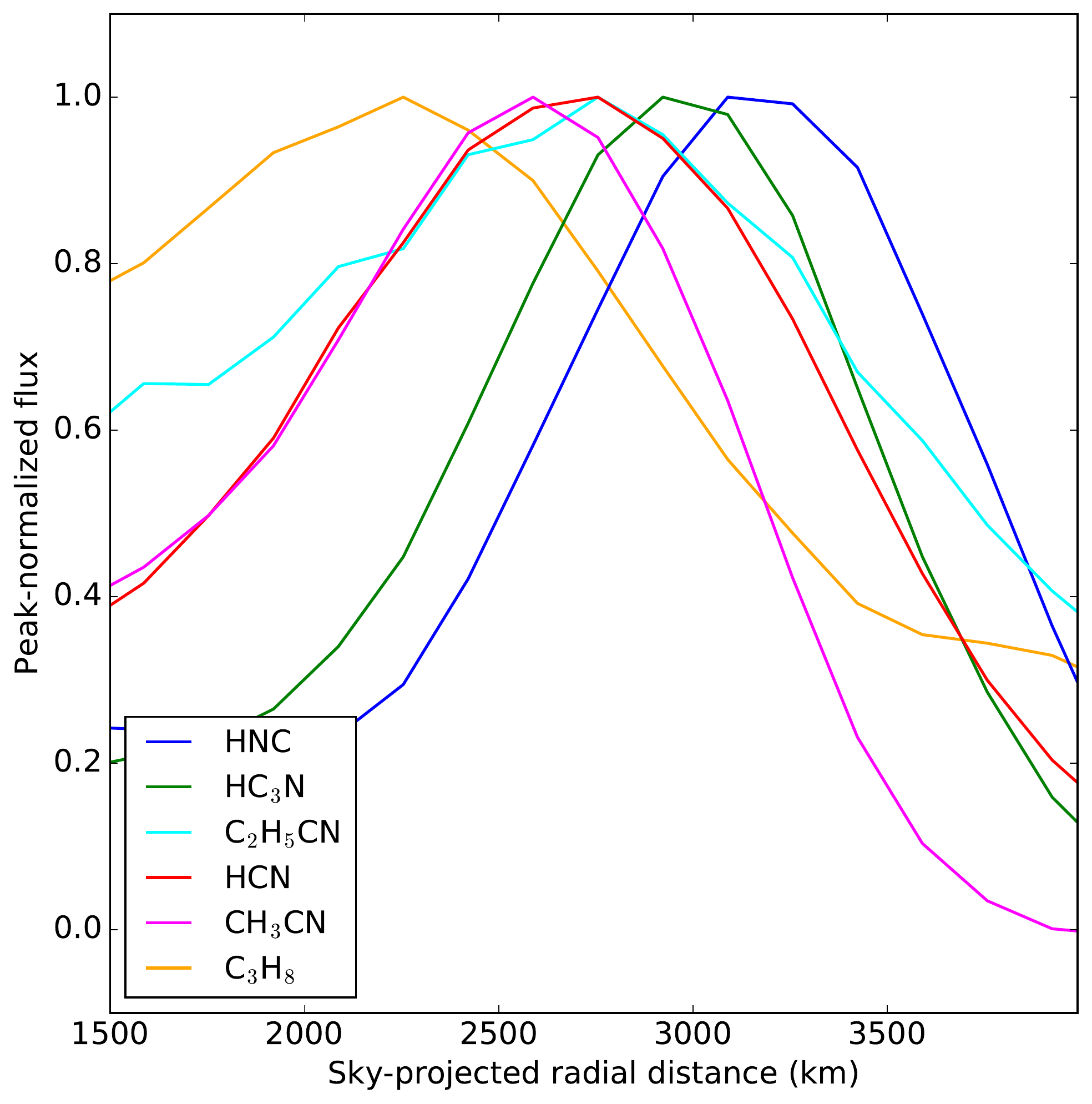}\vspace*{2mm}
\includegraphics[width=\columnwidth]{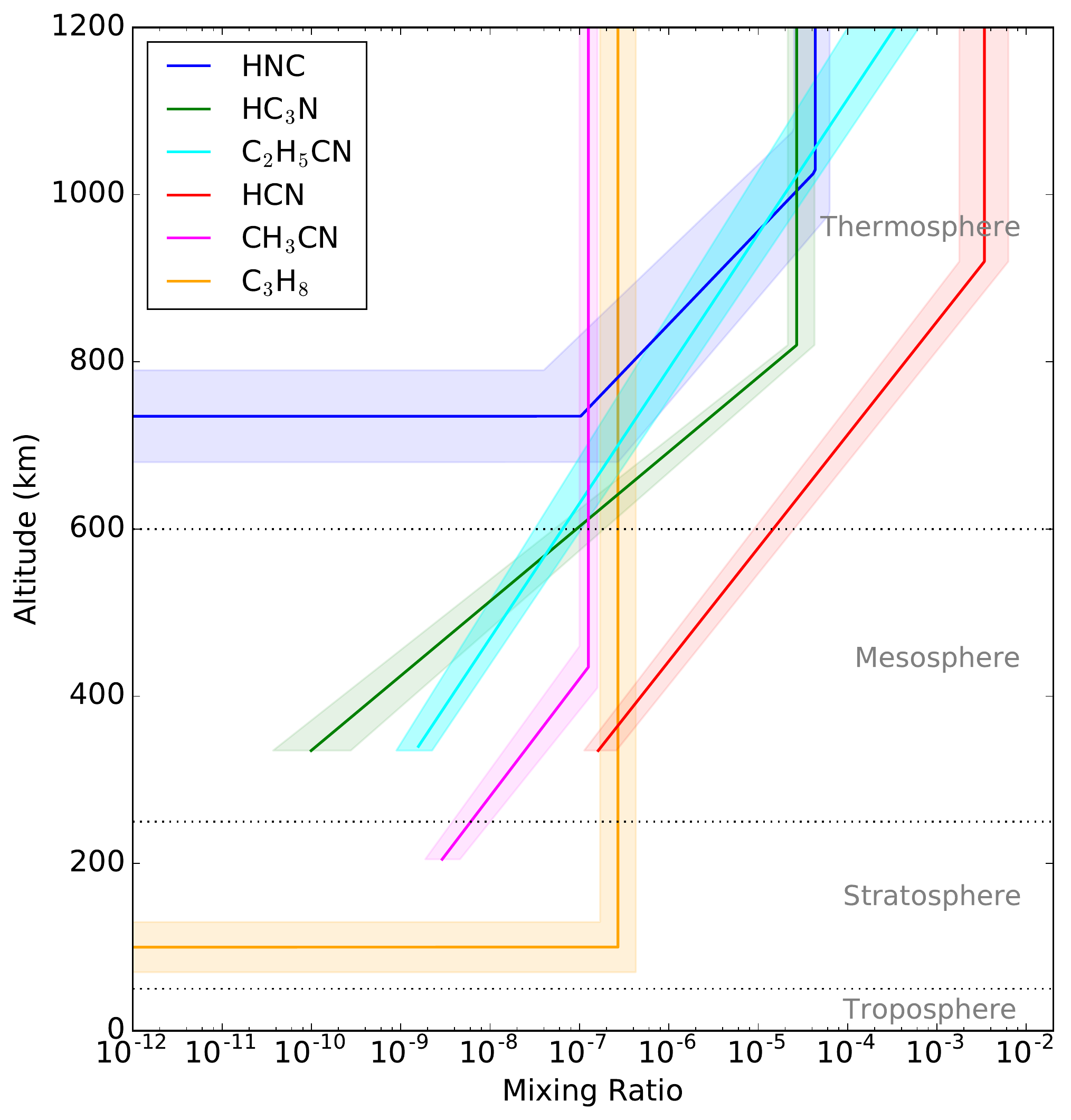}
\caption{Top panel: Peak-normalized equatorial limb flux profiles observed using ALMA (excluding the regions within $\pm45^{\circ}$ of Titan's north and south poles). Distances are with respect to the center of Titan's disk, projected in the plane of the sky. Bottom panel: Retrieved vertical abundance profiles {with $1\sigma$ error envelopes}.\label{fig:vmr}}
\end{figure}

The peak-normalized $S(r)$ profiles are shown in Figure \ref{fig:vmr} (top panel). Clear separations are present between the peak emission radii ($r_p$) for each species, {which were derived from the centroids of the $S(r)$ profiles, and are given in Table \ref{tab:alts}}. While the C$_3$H$_8$ and CH$_3$CN fluxes peak at $r_p=2400$~km and 2590~km, respectively (close to the edge of Titan's solid $r=2575$~km disk), \etcn, HCN, \cyano, and HNC peak well outside this radius, indicating flux contributions from high altitudes --- the very high HNC peak emission radius of {3140~km} is particularly surprising, and is consistent with a dominant thermospheric/ionospheric flux contribution for this gas. Unfortunately, an $S(r)$ profile ({and corresponding $r_p$ value}) could not be generated for \vycn\ due to a lack of flux from this molecule near the equator.

To further interpret the observed limb profiles, it is necessary to account for the line-of-sight averaging of different altitudes within each ALMA beam, as well as the atmospheric density (and temperature) variation with altitude. This was done by constructing a simple, spherically-symmetric molecular emission model for Titan. The density profile was taken from \citet{kra09}, and temperatures were retrieved from CIRS CH$_4$ limb observations at $-27^{\circ}$ latitude (observed during rev\#275), using the method of \citealt{tea07}. Molecular abundances as a function of altitude ($x(z)$) followed the simplest polygonal scheme possible while preserving a good fit to the observations: a logarithmic abundance gradient was adopted between altitudes $z_1$ and $z_2$, with zero abundance below $z_1$ and constant abundance above $z_2$. Spectral line fluxes were generated using parameters from the Cologne Database for Molecular Spectroscopy \citep{mul01}, under the assumption of optically thin emission. To reproduce the ALMA spatial response, the resulting images were passed through the ALMA simulator (using the {\tt Simobserve} and {\tt Clean} tasks), with the same observational and imaging parameters as described in Section \ref{sec:obs}. Synthetic image cubes were integrated over the same spectral ranges as the observations (Figures \ref{fig:maps1} and \ref{fig:maps2}), and radial flux profiles were extracted for comparison with observation.

The best-fitting $z_1$, $z_2$ and abundance gradient parameters were obtained through nonlinear least-squares optimization using the MPFIT routine \citep{mar12}, {and $1\sigma$ parameter errors were determined from the covariance matrix. The corresponding optimized limb flux models are shown for comparison with observations in Appendix A (Figure \ref{fig:append}), and retrieved abundance profiles are shown in Figure \ref{fig:vmr} (lower panel). Given the simplicity of our abundance parameterization scheme, the model limb flux profiles provide a surprisingly good fit for all species, with respect to the observational $1\sigma$ noise errors.}  

{To determine the altitude sensitivity of our abundance models, contribution functions (total flux as a function of altitude) were calculated within an ALMA beam centered at the respective $r_p$ value for each species.  The retrieved abundances are considered to be most reliable around the flux-weighted mean altitude ($\bar{z}_{\rm CF}$) of the contribution function, and these values are given in Table \ref{tab:alts}.}  Pressure-broadened line wings become significant {in models for} \cyano\ and \etcn\ sub-mm emission at altitudes below $\sim300$~km \citep{cor15,the19}, but such line wings are absent (or below the noise threshold) in our ALMA limb spectra, so the retrievals for these molecules contain little useful information below this altitude. We therefore truncate the abundance profiles for these species below 300~km in Figure \ref{fig:vmr}. We also truncate the HCN profile below 300~km due to the difficulty in reliably separating the wings of the $v_2=1e$ line from those of the ground-state HCN ($J=4-3$) line.  This region of the HCN ($J=4-3$) line wing originates predominantly from altitudes 100-200~km \citep{mol16}, and is strong enough to call into question the validity of the optically thin approximation for this molecule, which may therefore have also resulted in underestimating the overall HCN abundances. By virtue of its intrinsic line strength and high abundance at lower altitudes, pressure-broadened wings were clearly detected for \mecn, allowing the abundance for this species to be retrieved down to {$z\sim200$~km (determined using the \mecn\ contribution function of \citealt{the19}).}

Our ALMA limb observations provide the first well-constrained abundance retrievals for HNC and \etcn\ in the upper atmosphere. Prior (disk-averaged) observations of these molecules by \citet{mor11} and \citet{cor15} were only able to place lower limits (of 400~km and 200~km, respectively), on the altitudes from which the emission from these species originated. Our limb profile modeling shows that the detected HNC emission is confined to thermospheric altitudes above $730\pm60$~km, whereas \etcn\ follows a steep (logarithmic) abundance gradient from $(1.4\pm0.7)\times10^{-9}$ at 340~km to $(4\pm3)\times10^{-4}$ at 1200~km. {While the present article was undergoing peer review, the study of \citet{lel19} was published, also confirming that HNC is restricted to the thermosphere.}

\begin{table}
\centering
\caption{Peak ALMA limb emission radii ($r_p$), weighted mean contribution function altitudes ($\bar{z}_{\rm CF}$), and modeled lifetimes at 300~km ($L_{300}$)\label{tab:alts}}
\begin{tabular}{lllrr}
\hline\hline
Species & $r_p$ (km) & $\bar{z}_{\rm CF}$ (km) & $L_{300}$ (yr)$^a$ \\
\hline  
HCN      & 2670        & 530      & 350\\
HNC      & 3140        & 970    & 0.6\\
HC$_3$N     & 2930     & 760     & 2.6\\
C$_2$H$_3$CN   & ---   & ---    & 1.0\\
C$_2$H$_5$CN   & 2670  & 610     & 3.2$^b$\\
CH$_3$CN    & 2590     & 270     & 91\\
C$_3$H$_8$     & 2400  & 130   & 13\\ 
\hline
\end{tabular}
\parbox{\columnwidth}
{\footnotesize \vspace*{1mm} $^a$Lifetimes ($L_{300}$) are the chemical/dynamical lifetimes at 300~km from \citet{vui19}. $^b$The lifetime for C$_2$H$_5$CN is from a model including loss \emph{via} sticking to aerosols.}
\end{table}

\section{Comparison with Cassini CIRS}

Accounting for variations in atmospheric temperature and instrumental resolution, the primary morphology for HCN and HC$_3$N observed by CIRS is consistent with that revealed by our ALMA observations. Both Cassini and ALMA reveal that most of the HCN emission is concentrated in an atmospheric band (relatively uniform with latitude), at altitudes between 50-400~km.  The HC$_3$N emission is dominated by a higher-altitude, southern polar peak, confined to latitudes south of $-60^{\circ}$. The northern HC$_3$N peak detected by ALMA is barely visible in the Cassini nadir observations, which is consistent with the north-to-south column density ratio of $\approx1/7$ determined from the ALMA $v_7=1$ lines.  As explained in Section \ref{sec:specmap}, CIRS HCN radiances have a significant temperature dependency, which can explain the lack of a southern polar emission peak --- in fact, the CIRS HCN flux reaches a minimum at the south pole, where the ALMA flux reaches a maximum. 

\begin{figure}
\centering
\includegraphics[width=\columnwidth]{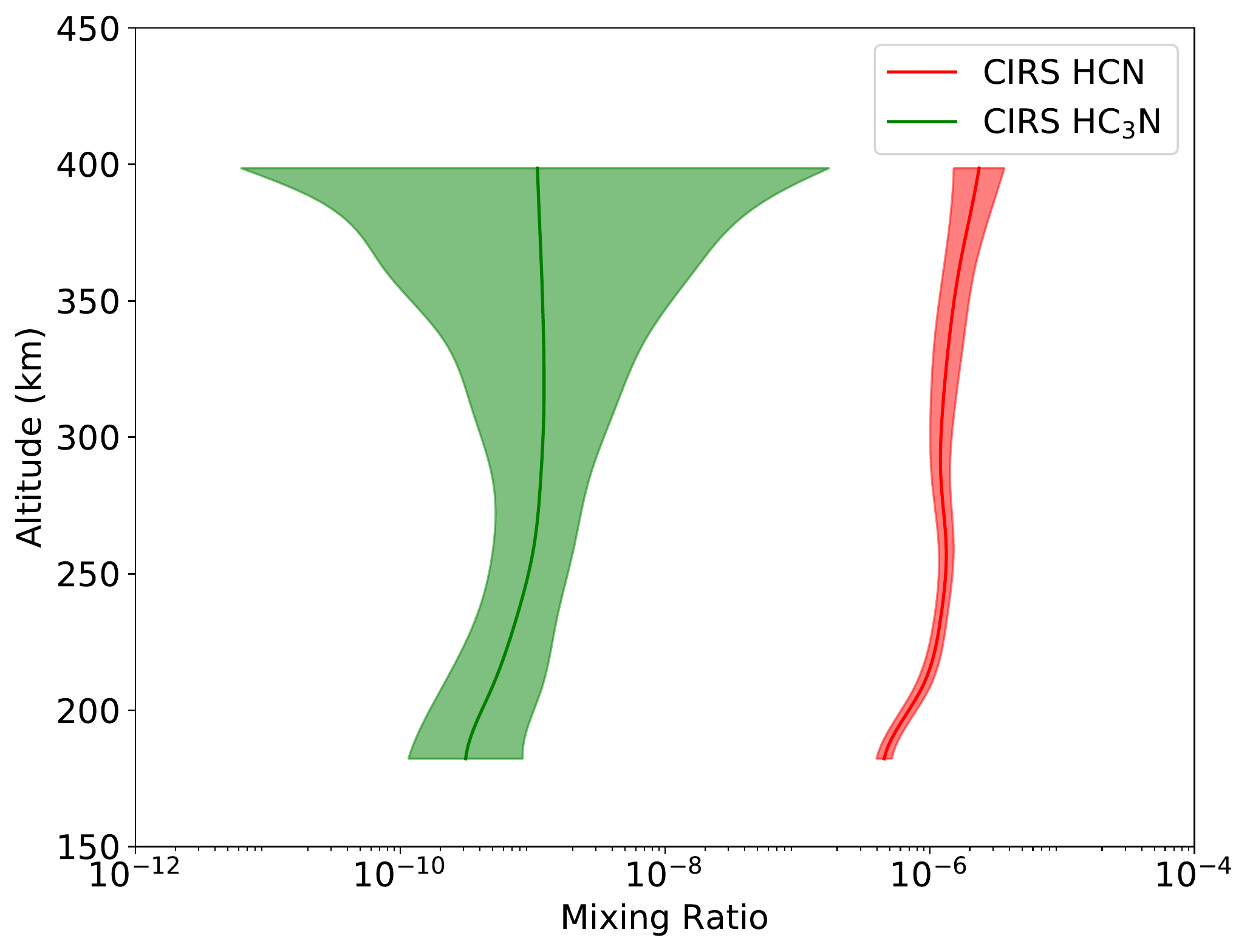}
\caption{Cassini CIRS limb abundances for HCN and HC$_3$N, retrieved at latitude $-11^{\circ}$.\label{fig:cirsvmr}}
\end{figure}

To enable a more direct comparison of the ALMA and Cassini observations, vertical profiles of temperature, HCN, and HC$_3$N abundance were retrieved from the CIRS data by modeling the limb radiances. The NEMESIS spectral inversion tool \citep{irw08} was used to fit the $-11^{\circ}$ limb observations from orbit \#261, following the method of \citet{tea07}. Prior to inverting for composition, a cubic b-spline curve was fitted to the measured radiance profiles at each wavenumber using a knot spacing of 100~km (see \citealt{tea07b} for further details). This improved the signal to noise by smoothing the radiance through multiple pixel measurements at similar but not identical tangent altitudes. The profiles obtained are reliable for 180-400~km; below 180~km the line opacity was too high to allow an accurate determination, and above 400~km the signal-to-noise was too low.

Retrieved equatorial limb abundances for HCN and HC$_3$N are shown in Figure \ref{fig:cirsvmr}. By comparison with Figure \ref{fig:vmr}, these are seen to be in reasonable agreement with the ALMA limb abundance profiles, especially when considering the error bars on the ALMA and CIRS retrievals, as well as the (unquantified) uncertainties in the ALMA limb abundances due to the low spatial resolution and simplified (polygonal) parameterization scheme used for those data.

\section{Discussion}

\subsection{Chemistry and dynamics of the thermosphere/ionosphere probed by HNC}

Our HNC observations reveal that this molecule is confined predominantly to the thermosphere, with an abundance of $(4.3\pm1.3)\times10^{-5}$ above 1030~km. The retrieved abundance profile (Figure \ref{fig:vmr}) is remarkably similar to the photochemical model results of \citet{vui19} (shown in Figure \ref{fig:vv}). The HNC model abundance profile rises steeply in the range 600-1000~km, and reaches a peak abundance $\sim2\times10^{-5}$ in the range 1050-1100~km. The {very good} agreement between model and observation for this species provides a strong validation for the chemical reaction scheme of \citet{vui19}, in which HNC is produced by electron recombination of the abundant HCNH$^+$ ion, and is rapidly converted back to the lower-energy HCN isomer (on a timescale of about a day), primarily by collisions with ionospheric H atoms. The short chemical lifetime means that HNC is destroyed before it can diffuse down through the atmosphere from its high-altitude production site, and is therefore confined to the thermosphere, with negligible abundance at lower altitudes. The rapid destruction of HNC makes it a potentially unique probe of very short-timescale dynamical processes in Titan's upper atmosphere.

{The observed asymmetry in HNC emission between Titan's eastern and western hemispheres can be explained as a result of the combined effects of photochemistry and atmospheric dynamics. The presence of mesospheric zonal winds (in the altitude range 250-450~km) was initially inferred on Titan based on the atmospheric structure derived from stellar occultation observations by \citet{hub93}. High-altitude (450-1000~km) superrotating winds were subsequently detected by \citet{mor05} using ground-based interferometry.} \citet{cui09} measured a strong (order of magnitude) diurnal variation in Titan's ionospheric HCNH$^+$ abundance using Cassini mass spectrometry, with the ion strongly depleted in the night hemisphere. They constructed a chemical/dynamical model for the ionosphere, which showed that east-west asymmetry in the distribution of HCNH$^+$ (and other ions) can be produced by a zonal wind that sweeps ion-rich material from the day side towards the eastern (dusk) terminator. The rapid conversion of HCNH$^+$ to HNC via electron recombination would then lead to an enhanced HNC abundance at the eastern limb. As it passes across the night side, collisional isomerization converts HNC back to HCN, resulting in subsequent emergence of HNC-depleted air in the west. {The detection by \citet{lel19} of superrotating zonal HNC winds extending up to the ionosphere} lends support to this theory, so in situ production of an asymmetric HCNH$^+$ distribution by electrons from Saturn's magnetosphere \citep[\eg][]{cra09} may not be required.

The HNC $J=4-3$ line flux has a relatively weak dependence on temperature, so a relatively large, 30-40 K hemispheric difference in rotational temperature would be required to explain the observed asymmetry. The atmospheric circulation model of \citet{mul00} predicted a diurnal temperature variation of 10-20~K (the precise value depends on solar activity), {at an altitude of 1300 km, corresponding to the top of Titan's thermosphere. However, this variation was found to fall with decreasing altitude, becoming negligible at the $\sim970$~km} altitude to which our HNC observations are most sensitive. Combined with the relatively low Solar activity around the time of our observations, diurnal temperature variations seem an unlikely explanation for the observed HNC asymmetry.

\subsection{High altitude nitrile chemistry}
\subsubsection{Inferred molecular lifetimes}

\begin{figure}
\centering
\includegraphics[width=\columnwidth]{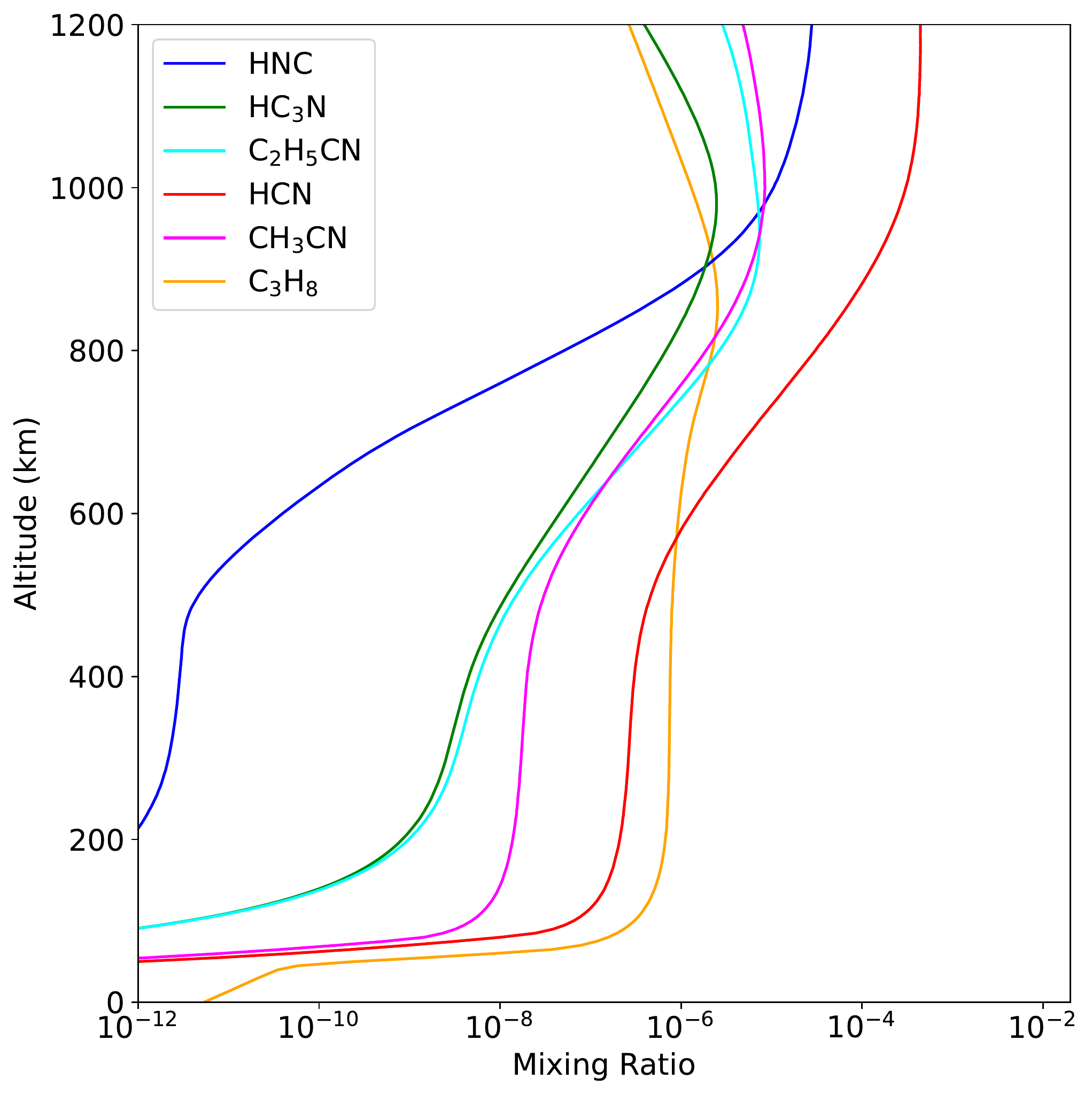}
\caption{Predicted equatorial abundances as a function of altitude, from the recent photochemical model of \citet{vui19}. The C$_2$H$_5$CN and HC$_3$N models include loss \emph{via} sticking to aerosols. \label{fig:vv}}
\end{figure}

The vertical abundance profiles retrieved for the remaining nitriles (HCN, CH$_3$CN, HC$_3$N and \etcn) are broadly consistent (within about an order of magnitude) with the results of recent chemical models \citep[\emph{e.g.}][]{kra09,loi15,wil16,dob16,vui19}, confirming the efficient production of organics through photochemistry in the thermosphere (above 600~km), initiated by Solar ultraviolet radiation. An abundance gradient develops as photochemical products mix downward towards the lower stratosphere, where the temperature becomes low enough for them to precipitate out. The steepness of the abundance gradient is affected by the chemical lifetime of each species, so from Figure \ref{fig:vmr}, we may infer that after HNC, HC$_3$N has the next shortest lifetime, followed by HCN, C$_2$H$_5$CN then CH$_3$CN. The overall trend is compatible with the lifetimes predicted at 300~km by \citet{vui19} (see Table \ref{tab:alts}), although it should be noted that their base model overestimated the stratospheric C$_2$H$_5$CN abundance, so the lifetime for this species is taken from their alternative model where C$_2$H$_5$CN is lost from the gas-phase \emph{via} sticking to stratospheric aerosol particles. The steep abundance profile for HCN also seems to be inconsistent with a long (350~yr) lifetime at 300 km. However, the HCN model lifetime varies strongly with altitude, and takes a value of only $\sim$ a few years in the lower thermosphere.

As a result of the coupling between the observed gas spatial distributions and Titan's seasonally varying global circulation, approximate lifetimes may also be derived from our ALMA emission maps, based on the rate of decay of abundance at the summer pole since the time of the last seasonal reversal.  We begin with the assumption that sufficient time has elapsed since the last seasonal reversal (following the 2009 equinox) that the molecular abundances at the southern (winter) pole are in a quasi steady-state, maintained by a balance between the source (Titan's global meridional circulation system) and sink (photochemical destruction). Then, neglecting temperature effects (which are expected to have only a small impact on the measured fluxes; see Section \ref{sec:specmap}), {the peak of the spectrally integrated flux in the southern hemisphere} ($F_S$) for a given species at the present epoch is taken to be representative of the {peak} flux that would have been present in the north at the time of last circulation reversal. This is supported by \citet{tea19} who suggest that north and south polar stratospheric abundances are closely comparable at similar seasonal phases, based on trends observed in CIRS nadir data across the entire Cassini mission. Using temporally-resolved molecular observations from CIRS, \citet{vin15} determined that the last circulation reversal occurred in 2011, so the peak north polar flux ($F_N$) has decayed to reach its present value over a period of 6 years. 

Assuming exponential decay of the form $dF_N/dt=F_Se^{-t/\tau}$, the following lifetimes $\tau$ were derived --- \cyano: 3.4~yr, \vycn: $<11.3$~yr, C$_2$H$_5$CN: 9.3~yr. These values are in reasonable agreement with the chemical/dynamical lifetimes ($L_{300}$) from the model of \citet{vui19} (see Table \ref{tab:alts}). We measure $F_N>F_S$ for CH$_3$CN, which  implies that the south polar abundance for this species has not yet reached steady state, so its lifetime cannot be derived using our method. However, the fact that the north and south polar fluxes differ indicates that the lifetime for this species is probably less than (or similar to) the 29.5~yr Saturnian year; if it were much longer, $F_N{\approx}F_S$ would be expected as a result of the repeated replenishment of chemically-enriched polar gases following successive winters. For HCN we find $F_N{\approx}F_S$, which implies a relatively long lifetime for the observed (polar) gases, consistent with the larger values of 60~yr and 350~yr at 300~km calculated by \citet{dob16} and \citet{vui19}, respectively.

These $\tau$ values represent an averaged measure of the molecular lifetime due to the combined influx/loss rates of molecules to/from the north polar region as a result of photolysis, chemical reactions, precipitation, diffusion, advection and winds, and may therefore not be considered as directly comparable with values from 1-dimensional chemical models. Additional caution is required for their interpretation due to the gross assumptions made in their derivation. In particular, the south polar flux at the present epoch may not be precisely representative of the northern value at the time of the 2011 circulation reversal, which may be the case if steady state was not reached in time, or in the case of seasonal climate asymmetry (\emph{i.e.}, if the atmosphere in Titan's northern winter is not a mirror image of the southern winter). Furthermore, the $27^{\circ}$ tilt of Titan's polar axis means that the south polar region was partially obscured from view at the time of our observations, so that $F_S$ provides an underestimate of the total flux. However, the majority of our observed flux is from altitudes above 200~km and latitudes $>-80^{\circ}$, all of which is within view, so geometrical obscuration is not expected to be a dominant source of uncertainty.

\subsubsection{Comparisons with prior observations and models}

Our ALMA limb abundance profiles for HCN, HC$_3$N and CH$_3$CN are in general, closely consistent with the prior equatorial retrievals from \citet{the19}, obtained by modeling ALMA spectra at lower spatial resolution. Our \etcn\ profile between 300 and 600 km is quite consistent with the disk-averaged gradient profile from \citet{cor15}. The thermospheric HCN profile also matches reasonably well with the Cassini VIMS limb observations of \citet{adr11}. It should be noted that the vertical abundance retrievals in our present study (based on spectrally-integrated limb fluxes) contain only limited information due to the relatively coarse ($>900$~km) spatial resolution of our data, and correspondingly simplified (polygonal) parameterization scheme for the abundances as a function of altitude. As such, our ALMA retrievals should be considered only approximate; full retrievals taking into account the detailed (pressure broadened) spectral line profiles will be presented in a future article. Nevertheless, some useful comparisons between our results and prior observations and models can still be made, particularly at higher altitudes (above 500~km) where previous (IR and radio) retrievals contain little useful information.

The retrieved HCN, HNC, \cyano, \etcn\ and C$_3$H$_8$ abundances at an altitude of 1000~km are close (within errors) to those predicted by the models of \citet{loi15}, \citet{wil16} and \citet{vui19}. Predicted vertical abundance profiles from \citet{vui19} are shown for comparison with our observations in Figure \ref{fig:vv}. For CH$_3$CN, these models over-estimate the ionospheric abundance by about an order of magnitude, which may indicate the presence of unaccounted-for (or underestimated) destruction pathway(s) for this molecule. It should be noted that although our HCN profile matches the model of \citet{vui19} reasonably well in the mesosphere and above, our retrieved HCN abundances are more uncertain towards lower altitudes due to the difficulty in separating the wings of the $J=4-3,\,v_2=1e$ line from those of the main ($J=4-3$) line.    

ALMA equatorial limb data show that HC$_3$N maintains a high abundance to lower altitudes than predicted by the \citet{loi15} and \citet{dob16} models, which fall too rapidly in the lower thermosphere (between 800-600~km). The models of \citet{wil16} and \citet{vui19} provide a better overall match for our HC$_3$N observations. Note, however, there is still some considerable uncertainty regarding the rate of loss of HC$_3$N \emph{via} sticking to aerosol particles in the \citet{vui19} model, so additional studies are needed to fully understand the chemistry of this species.

In summary, the reasonably good agreement between our nitrile observations and photochemical models for Titan's mesosphere and thermosphere indicates that the fundamental processes involved in the high-altitude production of nitriles are quite well understood. However, towards the stratosphere, the base model of \citet{vui19} does a relatively poorer job of matching the \cyano\ and \etcn\ observational data, which are over-predicted by at least an order of magnitude (a similar problem occurs in the lower stratosphere for the \citealt{loi15} and \citealt{dob16} models), indicating that important loss mechanisms for these molecules may be missing from chemical networks.  This issue may be addressed with the inclusion of sticking of these molecules to aerosol particles as an additional loss process in the stratosphere.

\section{Conclusions}

The ALMA observations presented here constitute a comprehensive, moderate-resolution mapping of Titan's most abundant atmospheric nitriles around the time of the 2017 solstice, coinciding with the end of the Cassini mission. This is the first time detailed, whole-hemisphere maps have been published for the complete set of nitriles accessible in the microwave/sub-mm band, and constitutes a unique resource for investigating their global distributions.  Each species shows a characteristic spatial distribution, resulting from the interplay between photochemical production, destruction and atmospheric dynamics. ALMA's unique capability to probe narrow molecular emission lines from high altitudes ($\sim300$-1000~km) in the mesosphere and thermosphere, in addition to pressure-broadened emission from lower, stratospheric altitudes ($\sim70$-300~km) provides complimentary data to that obtainable using infrared instruments such as Cassini CIRS and IRTF TEXES.

Our global maps for HC$_3$N, \vycn\ and \etcn\ reveal strong, compact emission peaks over the southern (winter) pole, indicative of rapid photochemical production (and destruction) at high altitudes, combined with transport towards southern latitudes by Titan's global meridional circulation system. Approximate lifetimes for these species are on the order of a few years, which matches the {stratospheric} chemical/dynamical lifetimes {(at $z=300$~km)} predicted by the latest chemical models for Titan's atmosphere, indicating that their dominant production and loss mechanisms are reasonably well understood. The HCN and \mecn\ maps show polar peaks with reduced contrast compared with the other species, {implying that they are relatively more well-mixed in latitude}, consistent with longer lifetimes for these gases (between about half and a few Saturnian years).

Limb flux profiles confirm that \mecn\ tends to be one of the more abundant nitriles at lower (stratospheric) altitudes, whereas the \cyano\ abundance in the stratosphere is low, and rises sharply with altitude throughout the mesosphere. The HNC vertical abundance profile is unique among the gases measured, as it does not become detectable until altitudes $\gtrsim700$~km in Titan's thermosphere. Thus, HNC provides a rare opportunity to probe chemical and physical processes occurring in Titan's thermosphere/ionosphere, using ground-based observations. 

The ALMA HNC map shows significant longitudinal asymmetry. This is explained as a result of the combined effects of zonal winds and diurnal variations in the ionospheric HCNH$^+$ abundance, which is theorized to be lower on the night side due to reduced solar insolation, giving rise to a reduced HNC abundance at the western (dawn) limb. The observed east-west asymmetry confirms the very rapid theorized production and loss rates for this species (on the order of a day at 1000~km altitude), making HNC a unique probe of short-timescale ionospheric processes.

\acknowledgements
This work was supported by the National Science Foundation under Grant No. AST-1616306, the NASA Astrobiology Institute (NAI), the NASA Solar System Observations program, and the UK Science and Technology Facilities Council (STFC). It makes use of ALMA data set ADS/JAO.ALMA\#2016.A.00014.S. ALMA is a partnership of ESO (representing its member states), NSF (USA), and NINS (Japan), together with NRC (Canada), and NSC and ASIAA (Taiwan), in cooperation with the Republic of Chile. The Joint ALMA Observatory is operated by ESO, AUI/NRAO, and NAOJ. The National Radio Astronomy Observatory is a facility of the National Science Foundation operated under cooperative agreement by Associated Universities, Inc.

%%%%%%%%%%%%%%%%%%%%%%%%%%%%%%%%%%%%%%%%%%%%%%%%%%



\bibliographystyle{aa}

%%%%%%%%%%%%%%%%%%%%%%%%%%%%%%%%%%%%%%%%%%%%%%%%%%

%%%%%%%%%%%%%%%%% APPENDICES %%%%%%%%%%%%%%%%%%%%%
\clearpage
\appendix

\section{Limb flux profile fits}
\label{sec:append}

\begin{figure*}[h!]
\centering
\includegraphics[width=0.32\textwidth]{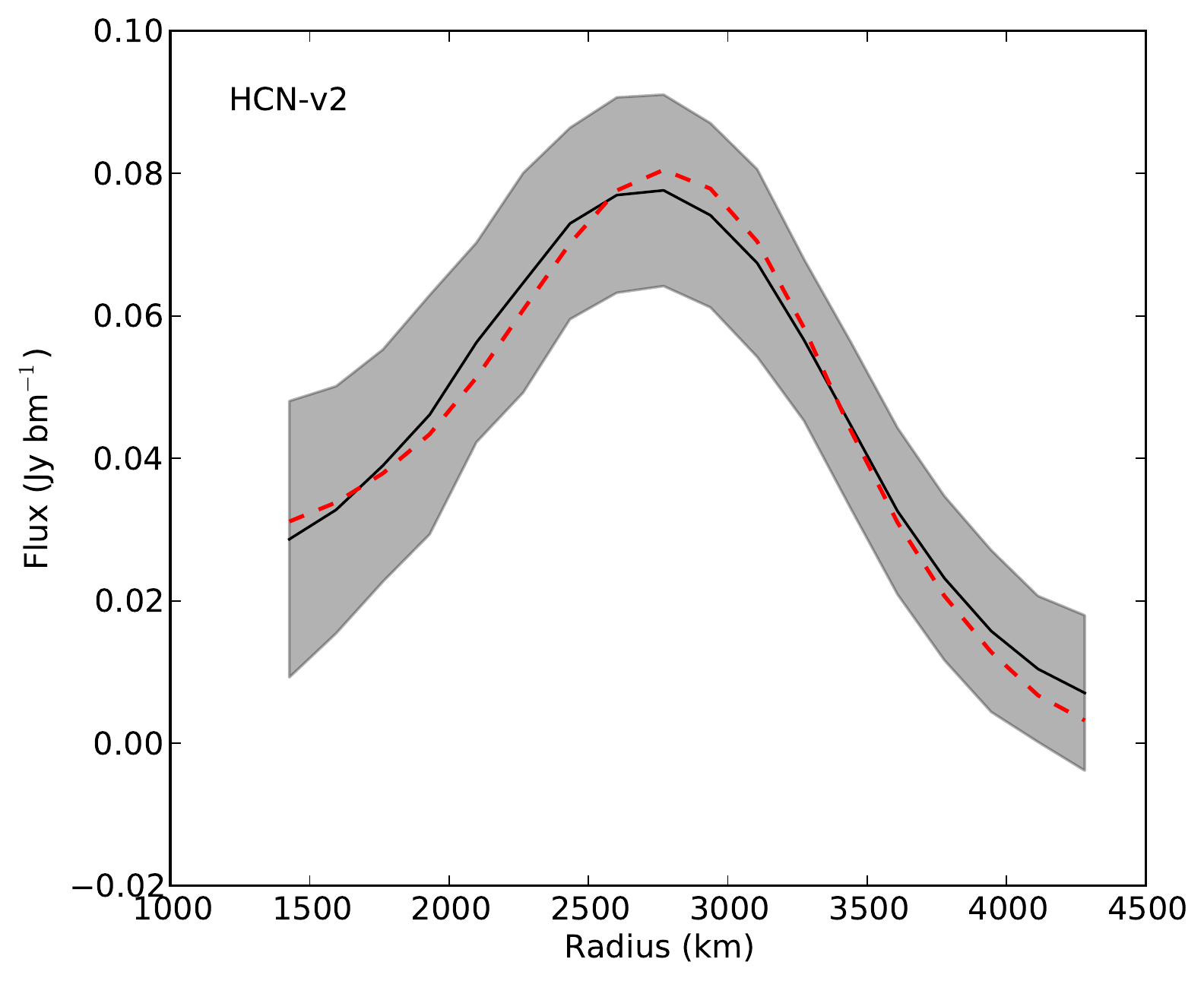}
\includegraphics[width=0.32\textwidth]{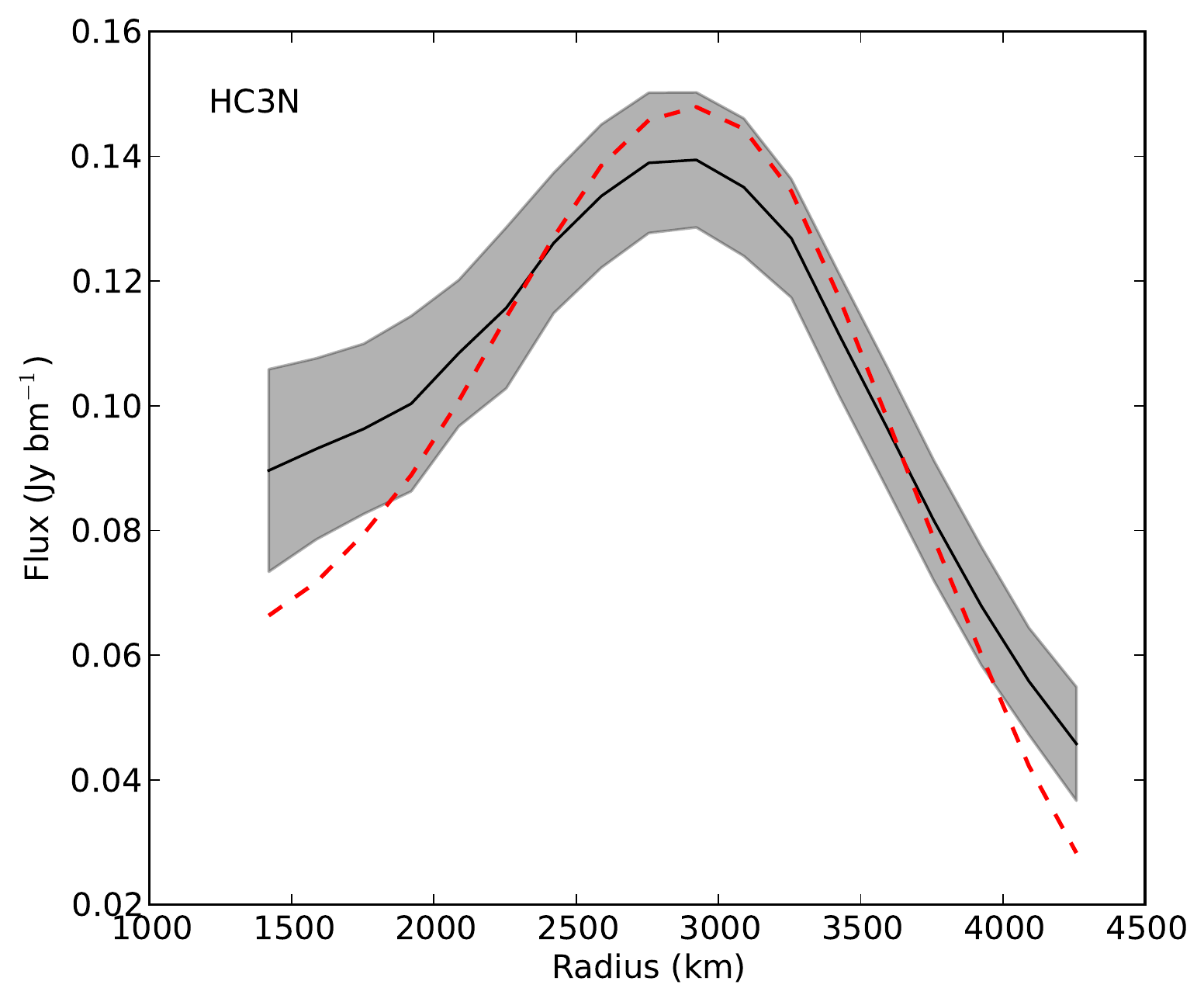}
\includegraphics[width=0.32\textwidth]{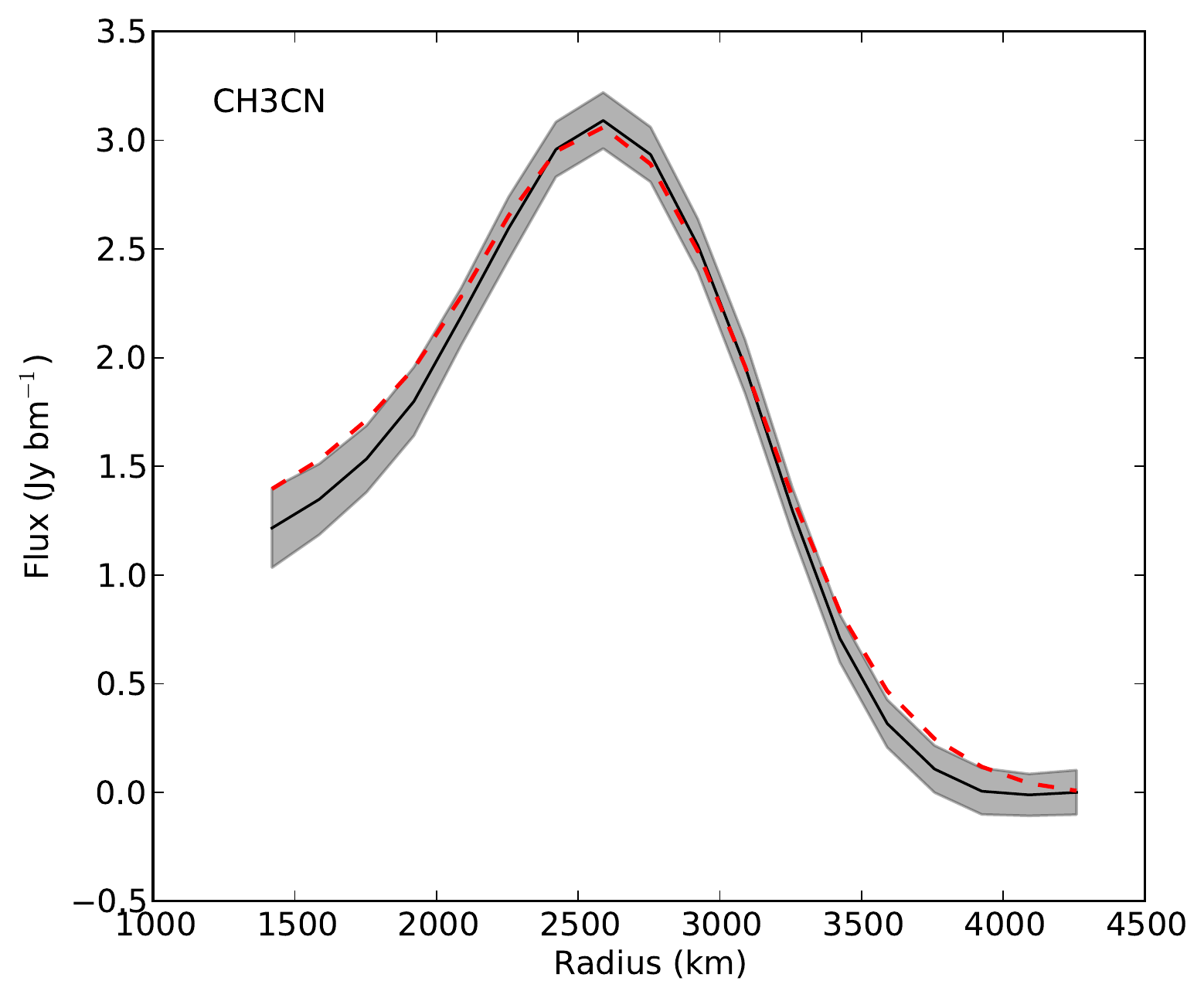}
\includegraphics[width=0.32\textwidth]{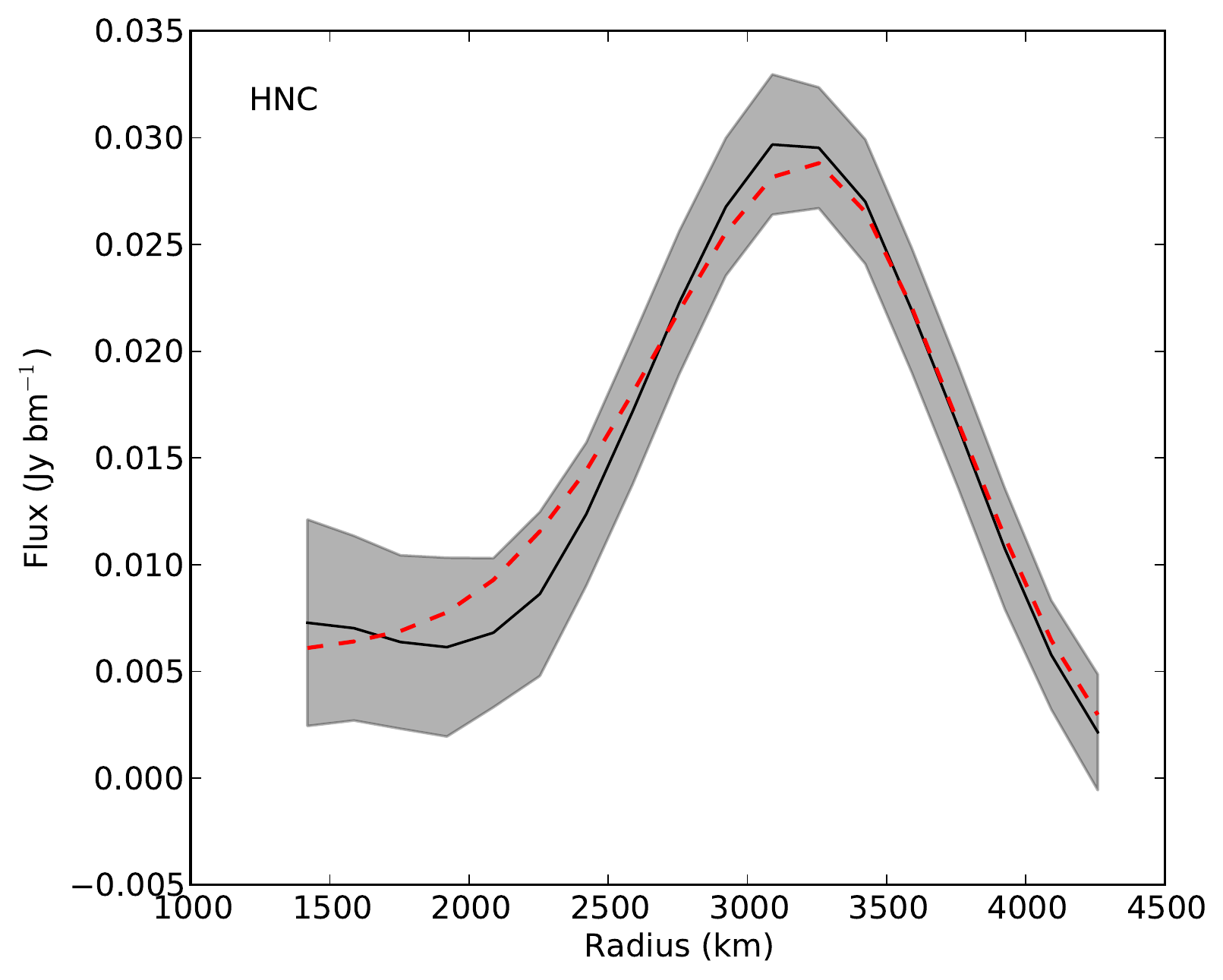}
\includegraphics[width=0.32\textwidth]{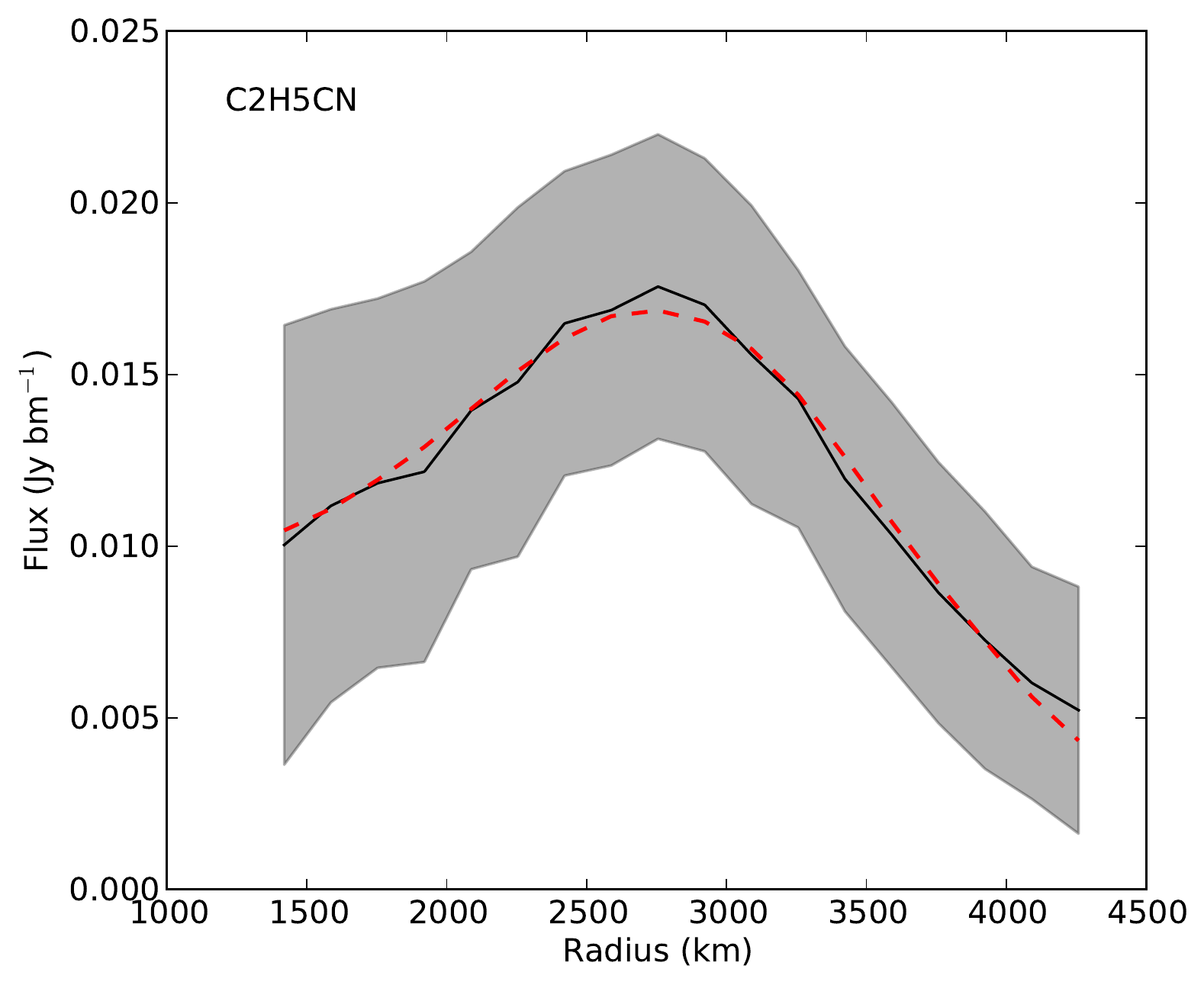}
\includegraphics[width=0.32\textwidth]{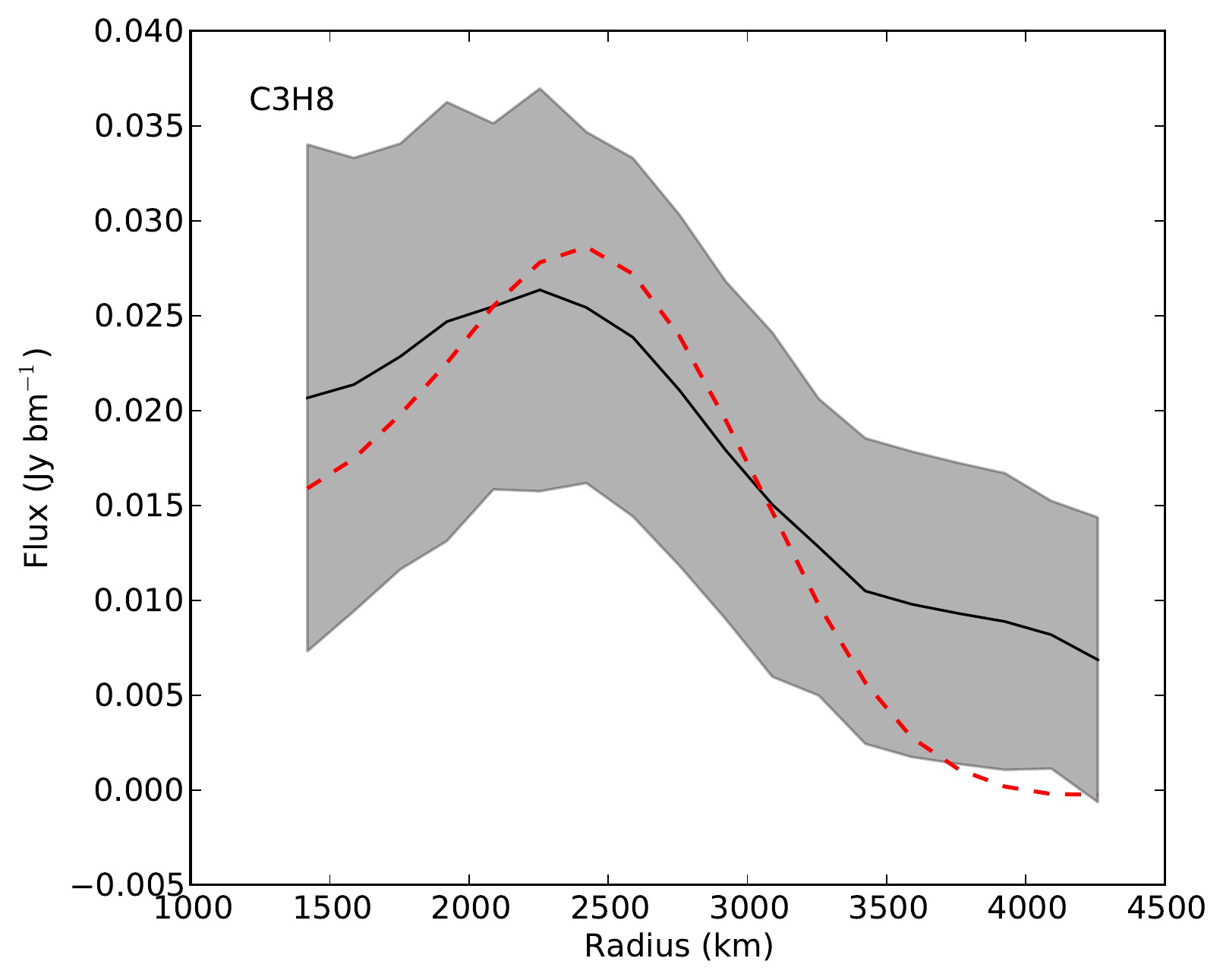}
\caption{Peak-normalized equatorial limb flux profiles observed using ALMA (excluding the regions within $\pm45^{\circ}$ of Titan's north and south poles). Grey envelopes indicate $1\sigma$ flux uncertainties, and the best-fitting model profiles are shown using red dashed curves. Radial distances are with respect to the center of Titan's disk, projected in the plane of the sky. \label{fig:append}}
\end{figure*}

{Least-squares fits to the equatorial fluxes as a function of (sky-projected) radius are shown in Figure \ref{fig:append}, using a simple, four-parameter model (two parameters for C$_3$H$_8$) for the abundance of each species ($x$) as a function of altitude ($z$). See Section \ref{sec:abund} for details of the modeling procedure. The corresponding optimized $x(z)$ profiles are shown in Figure \ref{fig:vmr}.}

%%%%%%%%%%%%%%%%%%%%%%%%%%%%%%%%%%%%%%%%%%%%%%%%%%

\end{document}